%% file: main.tex
\documentclass[journal=jctcce,manuscript=article,layout=traditional]{achemso}
\usepackage[hidelinks,hypertexnames=false]{hyperref}
\SectionNumbersOn
\usepackage[version=3]{mhchem} % Formula subscripts using \ce{}
\usepackage{amssymb}
\usepackage{xcolor}
\usepackage{subcaption}
\usepackage[frozencache]{minted}
\usepackage{tikz}
\usetikzlibrary{positioning,shapes,arrows,backgrounds,fit,calc,external,trees,tikzmark}
\usepackage{adjustbox}
\usepackage{dcolumn}
\usepackage{multirow}
\newcolumntype{d}[1]{D{.}{.}{#1}}

\usepackage{pdflscape}
\input{seaborn_colours}

\newcommand{\braopket}[3]{\left\langle #1 \right\vert #2 \left\vert #3 \right\rangle}
\newcommand{\ket}[1]{\left\vert #1 \right\rangle}

\tikzstyle{dummy} = []
\tikzstyle{line} = [draw, thick, -latex']
\tikzstyle{headless_line} = [draw, thick, -]
\tikzstyle{default}    = [rectangle, text centered, rounded corners, text=black, font=\sffamily\footnotesize, align=center]
\tikzstyle{default_text}    = [rectangle, text width=10cm, text=black,anchor=north west, font=\sffamily]
\tikzstyle{boxwhite} = [default, fill=white, rounded corners=0.1cm]
\tikzstyle{cp}    = [default, fill=seaborn_blue, text=white, text width=2.8cm, minimum height=0.7cm]
\tikzstyle{pw}    = [cp, fill=seaborn_green]
\tikzstyle{wannier90}    = [cp, fill=seaborn_cyan]
\tikzstyle{bespoke}    = [cp, fill=seaborn_magenta]
\tikzstyle{kcw}    = [cp, fill=seaborn_yellow]
\tikzstyle{w2k}    = [cp, fill=seaborn_magenta]
\tikzstyle{observable}    = [cp, fill=seaborn_red]
\tikzstyle{label} = [circle, text=white, fill=black, font=\sffamily\tiny\bfseries, minimum size=0.3cm, inner sep=1pt]
\tikzset{
  -|-/.style={
    to path={
      (\tikztostart) -| ($(\tikztostart)!#1!(\tikztotarget)$) |- (\tikztotarget)
      \tikztonodes
    }
  },
  -|-/.default=0.5,
  |-|/.style={
    to path={
      (\tikztostart) |- ($(\tikztostart)!#1!(\tikztotarget)$) -| (\tikztotarget)
      \tikztonodes
    }
  },
  |-|/.default=0.5,
}

\newlength{\myyshift}
\author{Edward Linscott}
\email{edward.linscott@epfl.ch}
\affiliation[THEOS]
{\small Theory and Simulation of Materials (THEOS), \'{E}cole Polytechnique F\'{e}d\'{e}rale de Lausanne, 1015 Lausanne, Switzerland}
\altaffiliation{Contributed equally to this work}
\author{Nicola Colonna}
\affiliation{Laboratory for Neutron Scattering and Imaging, Paul Scherrer Institut, 5232 Villigen, Switzerland}
\altaffiliation{Contributed equally to this work}
\alsoaffiliation[MARVEL]
{National Centre for Computational Design and Discovery of Novel Materials (MARVEL), \'{E}cole Polytechnique F\'{e}d\'{e}rale de Lausanne, 1015 Lausanne, Switzerland}
\author{Riccardo De Gennaro}
\affiliation[THEOS]
{\small Theory and Simulation of Materials (THEOS), \'{E}cole Polytechnique F\'{e}d\'{e}rale de Lausanne, 1015 Lausanne, Switzerland}
\author{Ngoc Linh Nguyen}
\affiliation{Faculty of Materials Science and Engineering, Phenikaa University, Hanoi 12116, Vietnam}
\alsoaffiliation{Phenikaa Research and Technology Institute (PRATI), A\&A Green Phoenix Group JSC, No. 167 Hoang Ngan, Trung Hoa, Cau Giay, Hanoi 11313, Vietnam}
\author{Giovanni Borghi}
\affiliation[THEOS]
{\small Theory and Simulation of Materials (THEOS), \'{E}cole Polytechnique F\'{e}d\'{e}rale de Lausanne, 1015 Lausanne, Switzerland}
\altaffiliation{Now at Liceo Manfredo Fanti, 41012 Carpi, Italy\vspace{-2ex}}
\author{Andrea Ferretti}
\affiliation{Centro S3, CNR--Istituto Nanoscienze, 41125 Modena, Italy}
\author{Ismaila Dabo}
\affiliation{Department of Materials Science and Engineering, Materials Research Institute, and Institutes of Energy and the Environment, The Pennsylvania State University, University Park, Pennsylvania 16802, USA}
\author{Nicola Marzari\vspace{-2ex}}
\email{nicola.marzari@epfl.ch}
\affiliation[THEOS]
{\small Theory and Simulation of Materials (THEOS), \'{E}cole Polytechnique F\'{e}d\'{e}rale de Lausanne, 1015 Lausanne, Switzerland}
\alsoaffiliation[MARVEL]
{National Centre for Computational Design and Discovery of Novel Materials (MARVEL), \'{E}cole Polytechnique F\'{e}d\'{e}rale de Lausanne, 1015 Lausanne, Switzerland}
\alsoaffiliation{Laboratory for Materials Simulations, Paul Scherrer Institut, 5232 Villigen, Switzerland}

\title[\texttt{koopmans}: predicting spectral properties with Koopmans functionals] {\large \texttt{koopmans}: an open-source package for accurately and efficiently predicting spectral properties with Koopmans functionals}

\begin{document}

\begin{tocentry}

   \includegraphics[width=3.25in]{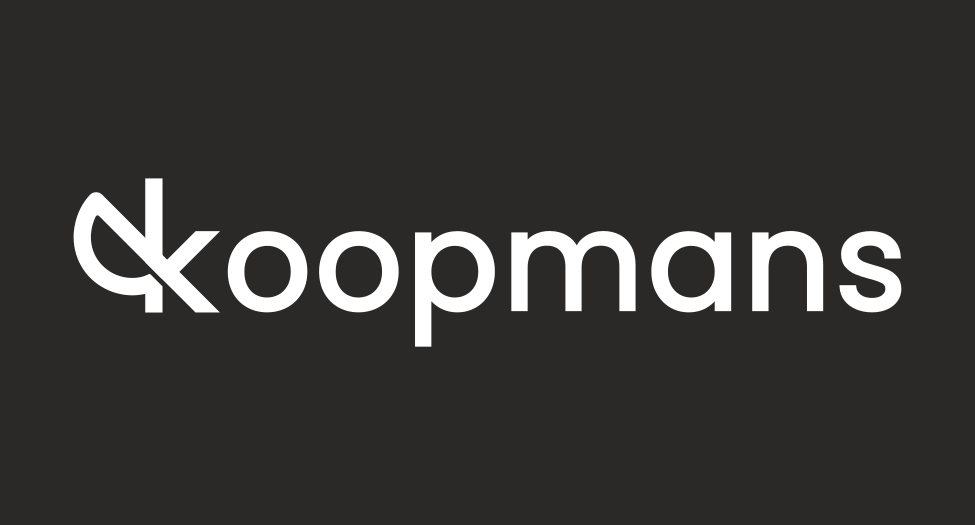}

   % Some journals require a graphical entry for the Table of Contents.
   % This should be laid out ``print ready'' so that the sizing of the

   % Inside the \texttt{tocentry} environment, the font used is Helvetica
   % 8\,pt, as required by \emph{Journal of the American Chemical
   %    Society}.

   % The surrounding frame is 9\,cm by 3.5\,cm, which is the maximum
   % permitted for  \emph{Journal of the American Chemical Society}
   % graphical table of content entries. The box will not resize if the
   % content is too big: instead it will overflow the edge of the box.

   % This box and the associated title will always be printed on a
   % separate page at the end of the document.

\end{tocentry}

\newpage
\begin{abstract}
   Over the past decade we have developed Koopmans functionals, a computationally efficient approach for predicting spectral properties with an orbital-density-dependent functional framework. These functionals impose a generalized piecewise linearity condition to the entire electronic manifold, ensuring that orbital energies match the corresponding electron removal/addition energy differences (in contrast to semi-local DFT, where a mismatch between the two lies at the heart of the band gap problem and, more generally, the unreliability of Kohn-Sham orbital energies). This strategy has proven to be very powerful, yielding molecular orbital energies and solid-state band structures with comparable accuracy to many-body perturbation theory but at greatly reduced computational cost while preserving a functional formulation. This paper reviews the theory of Koopmans functionals, discusses the algorithms necessary for their implementation, and introduces \texttt{koopmans}, an open-source package that contains all of the code and workflows needed to perform Koopmans functional calculations and obtain reliable spectral properties of molecules and materials.

   % \textcolor{red}{Decisions still to make
   %    \begin{itemize}
   %       \item Author list
   %       \item Is this paper is going to serve as a reference paper? If so, we might want to make the title more generic and pad out the theory section
   %       \item Do we want to present any results on
   %             \begin{itemize}
   %                \item computational scaling
   %                \item walltime comparisons
   %             \end{itemize}
   %    \end{itemize}
   %    %
   % }
\end{abstract}

\newpage

\section{Introduction}\label{sec_intro}
% Density functional theory
\noindent How can one accurately and efficiently predict spectral properties of molecules and materials \emph{ab initio}? Currently, the most accurate and popular approaches to compute charged excitation energies are Green's functions methods such as many-body perturbation theory (GW) \cite{Hedin1965,Aryasetiawan1998} or wavefunction methods such as quantum Monte Carlo \cite{Foulkes2001} and equation-of-motion coupled cluster \cite{Stanton1993} --- although for the latter, calculations for the solid state (rather than for molecules) are far from routine. Of these approaches, GW is computationally the least expensive, scaling as $\mathcal{O}(N^4)$, where $N$ is the number of electrons in the system.

Despite ongoing progress in the field of GW\cite{Golze2019}, performing these calculations is not straightforward. The aforementioned scaling of $\mathcal{O}(N^4)$ can still be an obstacle, and the calculations themselves can be challenging: they converge slowly with respect to the number of empty states included (which increases the importance of constructing transferable pseudopotentials that avoid ghost states\cite{vanSetten2018}), and there is a strong interdependence of the results on different calculation parameters, which makes achieving convergence challenging at best. This hampers routine applications of GW (especially in a high-throughput context, where the calculations must be unsupervised) \cite{Bonacci2023b}. Finally, while in principle GW and many-body perturbation theory are systematically improvable --- that is to say, by increasing the number of diagrams included in the calculations, the results should progressively converge to the correct answer (with GW outperforming GW\textsubscript{0} in turn outperforming G\textsubscript{0}W\textsubscript{0}) --- in practice this does not appear to hold\cite{Bruneval2021}.

Alternatively, one could try and calculate the energies of electronic excitations with density-functional theory (DFT) \cite{Hohenberg1964a,Kohn1965a}. DFT has proven to be a remarkably successful theory for predicting the ground-state properties of solids, surfaces, nanoparticles, and molecules \cite{Jones2015a,Marzari2021}. It is typically inexpensive, and these days such calculations are generally robust and can be treated as a ``black box". However, DFT is a theory of total energies, and while the Kohn-Sham auxiliary system is a powerful construct, the Kohn-Sham eigenvalues are not necessarily related to the energies of charged excitations (with the exception of the highest occupied molecular orbital, or HOMO, which is related to the exponential decay of the density \cite{Almbladh1985}). Nevertheless, these eigenvalues can bear qualitative or even quantitative resemblance to experimental quasiparticle energies, and it is common practice to interpret them as such, motivated by the fact that the Kohn-Sham potential is the best local and static approximation to the electronic self-energy~\cite{Casida1995}. Asides from this theoretical disconnect, problems also arise from the additional approximations inherent in exchange-correlation functionals. In the case of local and semi-local functionals, a key qualitative failure arises from the erroneous convex curvature in the total energy as a function of the total number of electrons in the system, which should instead be piecewise-linear \cite{Cohen2008}. This curvature explains in part the disagreement between first ionization potentials as calculated via total energy differences compared to Kohn-Sham eigenvalues.

Many strategies have emerged that attempt to restore the piecewise linearity of the energy functional --- the hope being that the resulting Kohn-Sham eigenvalues will yield accurate excitation energies. For example, DFT+\emph{U} imposes a penalty functional to a localized subspace that restores linearity in the energy with respect to the occupation of this subspace\cite{Anisimov1991a, Cococcioni2005a}. Similarly, hybrid functionals mix semi-local functionals with Hartree-Fock exchange (which happens to exhibit a concave curvature), which means that for a specific mixing fraction of the two functionals there will be an overall error cancellation \cite{Perdew1996, Marsman2008, Sai2011, Atalla2016}. Recent state-of-the-art approaches that employ curvature corrections to yield reliable quasiparticle energies include screened, range-separated, and dielectric-dependent hybrid functionals with tuned mixing or range-separation parameters\cite{Kronik2012,Brawand2016,Chen2018,Miceli2018,Skone2016a,Wing2021}, as well as the Koopmans-Wannier method of Wang and coworkers \cite{Ma2016} and the localized orbital scaling correction (LOSC) of Yang and coworkers \cite{Li2018,Yang2020,Mei2022}. Piecewise linearity is also central to ensemble density functional theory \cite{Valone1980,Baerends2022}. Even DM21, the recent machine-learned exchange-correlation functional created by Google DeepMind, was constructed around the idea of restoring piecewise linearity \cite{Kirkpatrick2021a}.

% While GGAs were included in the original paper by Kohn and Sham, they were not as reliable as the LDA \cite{?}. That changed when Perdew observed that the exchange-correlation hole $n_xc$ should integrate to $-1$. This condition is satisfied by the LDA by construction, but not by GGAs. When this condition was enforced, the performance of the GGAs greatly improved \cite{?}. It is worth noting that this advance came about by enforcing an exact condition, motivated by physical reasoning, and was not accompanied by an increase in computational demand.
% 
% Inspired by this lesson, Koopmans functionals seek to impose physically reasonable constraints on semi-local DFT functionals, in an attempt to improve their accuracy % without greatly improving computational cost --- I don't think this is a smart way of framing these functionals
% 
% Specifically, 

% Brief overview of Koopmans functionals
Starting in 2009, we have introduced and developed the concept of Koopmans functionals \cite{Dabo2009,Dabo2010,Dabo2013,Borghi2014,Ferretti2014,Borghi2015,Nguyen2015,Nguyen2016,Nguyen2018,Colonna2018,DeGennaro2022,Colonna2019,Schubert2023}. By imposing a generalized piecewise linearity condition and relating quasiparticle energies to total energy differences, these functionals address the above issues, and as a consequence they yield spectroscopic properties (such as molecular ionization potentials, electron affinities, solid-state band structures, and band-edge alignments) with comparable accuracy to state-of-the-art GW approaches, but at greatly reduced computational cost while preserving a functional formulation. This has all been implemented in \texttt{koopmans}, an open-source package that allows non-experts to perform their own Koopmans functional calculations, and which is built upon the popular \texttt{Quantum ESPRESSO} distribution. This paper provides an overview of the theory of Koopmans functionals (Section \ref{sec:theory}), describes the algorithms that enable their implementation in the \texttt{koopmans} package (Section~\ref{sec:implementation}), and demonstrates how these tools can be deployed to predict spectral properties using the examples of ozone, silicon, and zinc oxide (Section \ref{sec:examples}).

\section{Koopmans functionals}
\label{sec:theory}

\subsection{Fundamental concepts}
\label{sec:core_theory}

For a spectral theory, the orbital energies $\varepsilon_i$ should match the total energy differences corresponding to electron removal $E(N) - E_i(N-1)$ and addition $E_i(N+1) - E(N)$. This is trivially true for the exact Green's function, whose poles correspond directly to these total energy differences, but there is no such connection in Kohn-Sham DFT. The only exception to this is the HOMO, but even there the violation of piecewise linearity in density functional approximations leads to a mismatch between the HOMO eigenvalue with the corresponding total energy difference (i.e. the negative of the ionization potential).

Koopmans functionals restore this correspondence, by imposing the condition that the orbital energies $\varepsilon_i= \braopket{\varphi_i}{\hat H}{\varphi_i} = \frac{dE}{df_i}$ of orbitals $\varphi_i$ should be independent of that orbital's occupation $f_i$:
\begin{equation}
   \varepsilon_i = \text{constant with respect to}\ f_i
   \label{eqn:generalized_piecewise_linearity}
\end{equation}
It follows from Janak's theorem that this is equivalent to a ``generalized'' piecewise linearity condition where the total energy of the system is piecewise linear with respect to the change of occupation of \emph{any} orbital. This is a sufficient but not a necessary condition to fulfil the much more well-known piecewise linearity condition\cite{Perdew1982a}, which states that the total energy of the system is piecewise linear with respect to its \emph{total} number of electrons. In passing, we mention that eq.~\ref{eqn:generalized_piecewise_linearity} is reminiscent of a photoemission experiment, where an electron is removed from a Dyson orbital.

Imposing this condition will require a beyond-DFT approach, and is not simply a matter of correcting density functional approximations within a DFT framework. We can see that this must be the case by considering the exact density functional, for which the Koopmans corrections must be non-vanishing. (This is because while the negative of the HOMO energy for the exact density functional matches the ionization potential, there is no such guarantee for the other eigenenergies\cite{Almbladh1985}.)

The generalized piecewise linearity condition of eq.~\ref{eqn:generalized_piecewise_linearity} is imposed on a ``base" functional (here, approximate or exact DFT) by removing, orbital-by-orbital, the non-linear dependence of the energy $E$ on the orbital occupation $f_i$ and replacing it with a term that is linear in $f_i$:
\begin{equation}
   E^\text{Koopmans}
   = {E^\text{DFT}} + \sum_i \left[
      - \left(
      E^\mathrm{DFT}
      - \left.E^\mathrm{DFT}\right|_{f_i=0}
      \right)
      + f_i \eta_i
      \right]
   \label{eqn:Koopmans_functional_form_without_alpha}
\end{equation}
where $\left.E^\mathrm{DFT}\right|_{f_i=f}$ corresponds to the DFT energy of the $(N - 1 + f)$-electron system, with the occupancy of orbital $i$ constrained to be $f$. The first term in the square brackets removes the dependence of the total energy on $f$, and the second term replaces it with a term explicitly linear in $f$. This construction is reminiscent of the SIC functional of Ref.~\citenum{Heaton1987}, but here the correction is generalized to the entire electronic manifold.

Here, one must choose a suitable slope $\eta_i$ for this linear term; one option is to use the energy difference between fully-occupied and empty orbitals
\begin{equation}
   \eta^\mathrm{KI}_i = \left.E^\mathrm{DFT}\right|_{f_i = 1} - \left.E^\mathrm{DFT}\right|_{f_i = 0}
   \label{eqn:eta_KI}
\end{equation}
giving rise to the Koopmans integer (KI) functional. Note that this formulation provides Koopmans functionals with meaningful eigenvalues, because they now correspond to total energy differences, which in the scope of DFT are formally meaningful and much more reliable than Kohn-Sham eigenvalues. It can be seen from eqs.~\ref{eqn:Koopmans_functional_form_without_alpha} and \ref{eqn:eta_KI} that the KI functional gives, at integer occupations, the same total energy as the base functional, but has different derivatives and hence yields different spectral properties. (This will be discussed further in Section~\ref{ki}).

Equations~\ref{eqn:Koopmans_functional_form_without_alpha} and \ref{eqn:eta_KI} are difficult to evaluate unless we only consider the explicit dependence of the DFT energy on the orbital occupancies, neglecting the implicit dependence of the orbitals $\varphi_i(\mathbf{r})$ on their own occupation $f_i$, in which case
\begin{equation}
   \left.E^\mathrm{DFT}\right|_{f_i=f}
   = E^\mathrm{DFT}[\rho - \rho_i + f n_i]
\end{equation}
where $n_i(\mathbf{r}) = |\varphi_i(\mathbf{r})|^2$ is the density of orbital $i$ and $\rho_i(\mathbf{r}) = f_i |\varphi_i(\mathbf{r})|^2 = f_i n_i(\mathbf{r})$ is the occupancy-weighted density of orbital $i$. Orbital relaxation --- or, equivalently, screening --- is instead accounted for \emph{post hoc} by scaling the unscreened correction by a scalar coefficient $\alpha_i$. Crucially, these coefficients can be calculated \emph{ab initio} at the level of DFT via linear response or total energy differences \cite{Nguyen2018,Colonna2018}. This brings us, finally, to the Koopmans energy functional:
\begin{equation}
   E^\text{Koopmans} [\rho,
      {\{\rho_i\}}]
   = {E^\text{DFT}[\rho]} + \sum_i \alpha_i \left[
      - \left(
      E^\mathrm{DFT}[\rho] - E^\mathrm{DFT}[\rho - \rho_i]
      \right)
      + f_i \eta_i
      \right]
   \label{eqn:Koopmans_functional_form}
\end{equation}
\begin{figure}
   \includegraphics[width=3.5in]{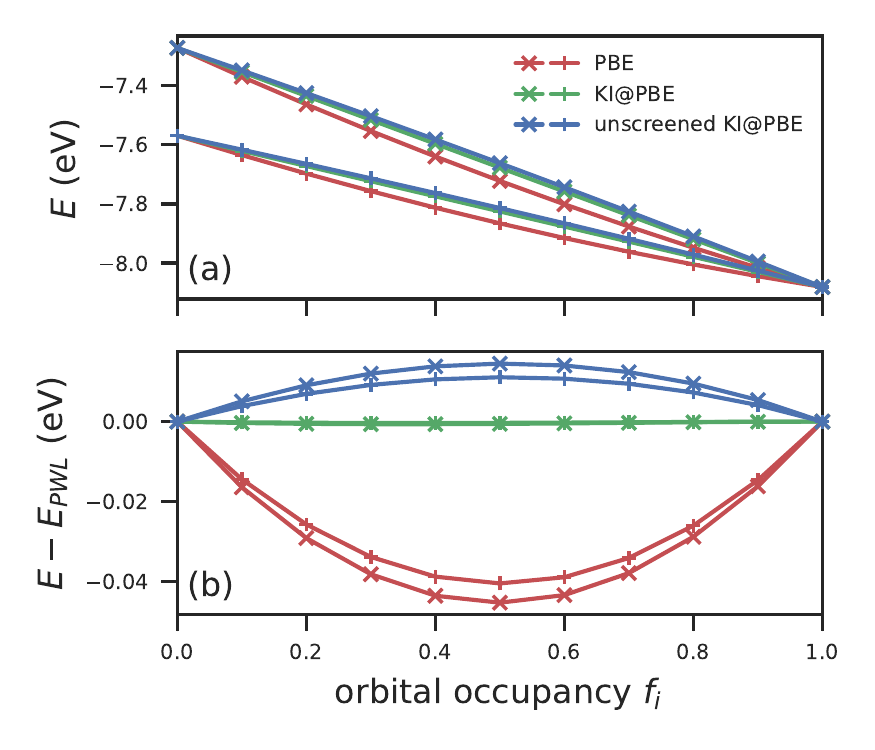}
   \caption{The total energy $E$ of a CH\textsubscript{4} molecule as a function of the occupancy of the $1a_1$ molecular orbital (crosses) and one of the $1t_2$ molecular orbitals (pluses). The absolute total energy is shown in (a) while the deviation from piecewise linearity is shown in (b). For both orbitals, PBE gives a total energy that is erroneously convex, while the KI correction successfully linearizes the total energy. Screening is key to this success; in its absence, the KI correction overcorrects the PBE base functional and yields a concave energy curve. Note that each orbital $\varphi_i$ is obtained from the charge-neutral system ($f_i = 1$) and is frozen throughout (while all others are relaxed). If that orbital was not frozen, then as $f_i \rightarrow 0$ the orbital would always morph into the LUMO of the $N-1$-electron system and both sets of curves would be identical.}
   \label{fig:pwl}
\end{figure}
In Figure~\ref{fig:pwl} we show the efficacy of this linearizing correction when applied to two orbitals in methane. The full derivation of eq.~\ref{eqn:Koopmans_functional_form} can be found in Supporting Information \ref{sec:derivation_of_functional_form}. This functional is actually very different from semi-local DFT functionals; this will be elaborated upon in the following sections.

\subsection{Orbital-density dependence}
\label{sec:odd}
The one important distinction that is worth making immediately is that Koopmans functionals are not density functionals, but \emph{orbital-density-dependent} (ODD) functionals. This is because they are dependent on the individual orbital densities $\{\rho_i\}$ and not just the total electronic density $\rho$. A direct consequence of this is that Koopmans functionals --- much like other ODD functionals such as the Perdew-Zunger self-interaction correction (PZSIC) --- are no more invariant under unitary transformation of the occupied manifold, and their minimization requires extra care. The variation of $E^{\rm Koopmans}$ in eq.~\ref{eqn:Koopmans_functional_form} with respect to an arbitrary change of each orbital $\varphi_i$ (density $\rho_i$) leads to the Euler-Lagrange equations
\begin{equation}
   h^{\rm DFT}|\varphi_i \rangle + v_i^{\rm ODD}|\varphi_i \rangle = \sum_j \Lambda_{ji} |\varphi_j \rangle
   \label{eqn:euler-lagrange}
\end{equation}
where $h^{\rm DFT}(\mathbf{r}) = \frac{\delta E^{\rm DFT}}{\delta \rho(\mathbf{r})}$ is the Hamiltonian of the underlying DFT energy functional, $v_i^{\rm ODD}(\mathbf{r})$ is the orbital-density-dependent potential associated with the orbital $\varphi_i$, and $\Lambda_{ji}$ is the matrix of Lagrangian multipliers enforcing orthonormality constraints. Because of the ODD contribution, within the space spanned by the orbitals $\{ \varphi_i \}$) the energy is representation-dependent and a proper minimization of the functional requires its variation with respect to infinitesimal unitary transformations among the occupied orbitals to vanish \cite{Pederson1984, Goedecker1997, Borghi2015}, leading to the Pederson condition \cite{Pederson1984}
\begin{equation}
   \langle \varphi_i | h_i | \varphi_j \rangle = \langle \varphi_i | h_j | \varphi_j \rangle.
   \label{eqn:Pederson_condition}
\end{equation}
The self-consistent solution of eqs.~\ref{eqn:euler-lagrange} and \ref{eqn:Pederson_condition} define the proper minimum of the Koopmans functionals, and the minimizing orbitals are known as the \emph{variational} orbitals. The implementation of this minimization procedure will be discussed later in Section~\ref{sec:dscf_implementation}.

\begin{figure}[t]
   \centering
   \begin{subfigure}[b]{0.4\columnwidth}
      \includegraphics[width=\columnwidth]{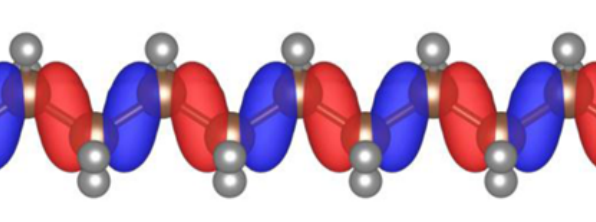}
      \caption{a canonical orbital}
      \label{fig:canonical_orbital}
   \end{subfigure}
   \hspace{0.1\columnwidth}
   \begin{subfigure}[b]{0.4\columnwidth}
      \includegraphics[width=\columnwidth]{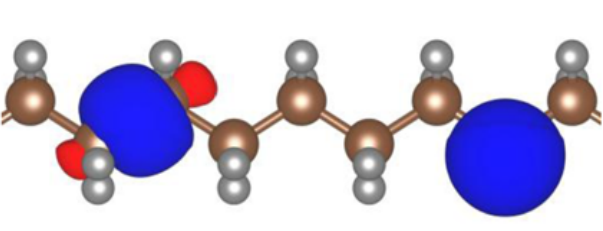}
      \caption{two variational orbitals}
      \label{fig:variational_orbital}
   \end{subfigure}
   \caption{Example canonical and variational orbitals of polyethylene. Figure adapted from Ref.~\citenum{Nguyen2018}.}
\end{figure}

At the minimum, as a consequence of eq.~\ref{eqn:Pederson_condition}, the $\Lambda$ matrix becomes Hermitian and can be diagonalized allowing us to define a set of \emph{canonical} orbitals and energies. This mirrors the definition of canonical orbitals and energy in Hartree-Fock theory where, among all the equivalent sets of orbitals (those related by a unitary transformations) that minimize the functional, the canonical orbitals are recognized as those that also make the energy functional stationary when a fraction of electron is added to or removed from the system, thus qualifying these as electron addition/removal energies. This also applies to ODD functionals, as discussed in detail in Ref.~\citenum{Pederson1985} for the case of PZSIC. Moreover, canonical orbitals typically display the symmetry of the Hamiltonian operator (e.g. are Bloch states in periodic systems~\cite{DeGennaro2022} as shown in Fig.~\ref{fig:canonical_orbital}) and, in analogy to exact DFT, the energy of the highest occupied canonical orbitals has been numerically shown to determine the asymptotic decay of the ground-state charge density~\cite{Stengel2008}. For all these reasons, the canonical orbitals and the corresponding eigenvalues are usually interpreted as Dyson orbitals and quasiparticle energies. Nevertheless, it is important to stress that the reliability of canonical energies (and their correspondence with total energy differences) is not directly imposed by the Koopmans correction, but instead is inherited via the variational orbitals. That is to say: the Koopmans corrections are applied to the variational orbitals, and thus the Koopmans functional is linear with respect to the occupancy of variational orbitals. The canonical orbitals are composed of some linear combination of variational orbitals, and their energies (i.e. the quasiparticle energies) are subject to a weighted combination of corrective potentials arising from their constituent variational orbitals.

Given their central role in the theory, it is important to discuss the key features of variational orbitals. In contrast to canonical orbitals, variational orbitals are typically very localized in space (see Fig.~\ref{fig:variational_orbital}). As was recognized long ago\cite{Pederson1984}, eq.~\ref{eqn:Pederson_condition} is a localization condition that, once satisfied, leads to orbitals that resemble Boys orbitals in molecules or, equivalently, maximally localized Wannier functions in periodic systems\cite{Marzari2012}. The localization of the variational orbitals is a common feature of ODD functionals and a key property for Koopmans functionals, in particular when it comes to dealing with periodic systems. By applying Koopmans corrections to a set of localized orbitals, the corrections are well-defined and non-vanishing for both small molecules, infinite bulk systems, and everything in between, preserving size-consistency \cite{Nguyen2018}. Contrast this to if we were to apply the corrections to the canonical orbitals, in which case they would become ill-defined in the bulk limit. In order to understand why this is the case, it is useful to return to the connection between the Koopmans construction and the $\Delta$SCF approach. In a nutshell, the ultimate effect of the Koopmans correction is to revert the wrong eigenvalue from the underlying (approximate) density functional into a total energy difference ($\Delta$SCF) between the neutral system and the system with plus or minus one electron evaluated using the same density functional. This means that the success of the approach relies on the quality of the $\Delta$SCF value at the approximate DFT level. It is well known that evaluating this total energy difference when removing an electron from a completely delocalized state reduces to the derivative of the total energy with respect to the particle number \cite{Mori-Sanchez2008,Kraisler2014a,Vlcek2015}, which, for a local or semilocal density-functional approximation, is the negative of the KS-DFT eigenvalue. This means that for a standard density functional in the thermodynamic limit there is no difference between the $\Delta$SCF and the KS eigenvalues and as a consequence the Koopmans corrections vanish. To overcome this issue, two routes are possible: either improving the base functional in such a way to have improved $\Delta$SCF energies in the most general case, or retaining the simplicity of local and semi-local density-functionals and working in a localized representation of the orbitals~\cite{Chan2010,Ma2016}. Indeed, the total energy differences of approximate density functionals also become accurate when computed on localized orbitals (e.g. typically, semi-local $\Delta$SCF calculations accurately predict localized defect levels relative to the average electrostatic potential \cite{Komsa2011}). Thus, by applying the Koopmans corrections to the variational orbitals (and not the canonical orbitals), the Koopmans corrections are well-defined and non-vanishing also in the bulk limit, and yield accurate band structures compared to experiment. See Ref.~\citenum{Nguyen2018} for more details.

%Each orbital is subject to a different potential, and when we minimize a Koopmans functional we must minimize the total energy with respect to the entire set of orbitals, as opposed to just the total density.
%One characteristic of ODD functional theories (ODDFTs) is that they can give rise to \emph{two} sets of orbitals that we must be careful to distinguish. The set of orbitals that minimize the total energy are called \emph{variational} orbitals. These variational orbitals tend to be very localized (see Figure~\ref{fig:variational_orbital}).
%
%
%Having minimized the Koopmans functional one can then construct the Hamiltonian. If we then diagonalize this Hamiltonian we would obtain the so-called \emph{canonical} orbitals. 
%Moreover, in contrast to the variational orbitals, the canonical orbitals much more closely resemble the Kohn-Sham orbitals of DFT and Dyson orbital of Green's function theory (see Figure~\ref{fig:canonical_orbital}). In a solid, they satisfy Bloch's theorem \cite{DeGennaro2022}. 
%In a DFT framework, the energy of the system is the same if we evaluate it using the canonical orbitals or the variational orbitals, because the energy depends only on the total density. However, in an ODDFT, this is not the case: the total energy is not invariant with respect to unitary rotations of a given set of orbitals, and thus the ODD energy differs depending on which set of orbitals it is evaluated.

Moving from a DFT framework to an ODDFT framework may appear like an unnecessary complication. This is not the case: ODDFTs are a very natural way to generalize a static functional theory like DFT to predict spectral information. Ultimately, the spectral properties of a many-body electronic system are exactly described by its non-local and dynamic self-energy. The exact Kohn-Sham potential is the best local and approximation to this self-energy \cite{Casida1995}. If we instead consider local but dynamic approximations, one enters into the domain of spectral functional theories, where the exact spectral functional predicts exactly the spectral density $\rho(\mathbf{r}, \omega)$~\cite{Gatti2007}. ODDFTs can be interpreted as energy-discretized spectral functional theories~\cite{Ferretti2014}, so as such an ODDFT framework is a sensible choice when attempting to predict spectral properties.

%MOVED ABOVE NiCo
%Secondly, it is very apt that the ODD formalism naturally gives rise to a set of localized orbitals when it comes to dealing with periodic systems. By applying Koopmans corrections to a set of localized orbitals, the corrections are well-defined and non-vanishing for both small molecules, infinite bulk systems, and everything in between. Contrast this to if we were to apply the corrections to the canonical orbitals, in which case they would become ill-defined in the bulk limit. For a more detailed discussion on this topic, see Supporting Information~\ref{sec:the_bulk_limit}.

%MOVED ABOVE NiCo
%Nevertheless, the canonical orbitals and the corresponding eigenvalues are always interpreted as Dyson orbitals and quasiparticle energies. Note that this means that the reliability of quasiparticle energies (and their correspondence with total energy differences) is not directly imposed, but instead is inherited via the variational orbitals. That is to say: the Koopmans corrections are applied to the variational orbitals, and thus the Koopmans functional is linear with respect to the occupancy of variational orbitals. The canonical orbitals are composed of some linear combination of variational orbitals, and their energies (i.e. the quasiparticle energies) will be subject to some weighted combination of corrective potentials arising from their constituent variational orbitals.

\subsection{Accounting for screening effects}
\label{sec:calculating_alphas}

As discussed earlier in Section~\ref{sec:core_theory}, we account for orbital relaxation \emph{post hoc} via screening parameters $\{\alpha_i\}$ and we can calculate these parameters \emph{ab initio}. But how?

The crucial point is that we would like the total energy to be piecewise linear: that is, we would like orbital energies (specifically, the expectation value of the Hamiltonian on a given variational orbital) to match the corresponding total energy differences when adding/removing an electron from this orbital, without the frozen-orbital assumption that we made earlier. Specifically, we would like $\lambda_{ii}(\alpha, f) = \Delta E^\text{Koopmans}_i$, where
\begin{equation}
   \lambda_{ii}(\alpha, f) = \left.\braopket{\varphi_i}{\hat h^\mathrm{DFT} + \alpha \hat v^\mathrm{Koopmans}}{\varphi_i}\right\vert_{f_i=f}
   = \left.\frac{\partial E^\text{Koopmans}}{\partial f_i}\right\vert_{f_i = f}
   \label{eqn:lambda_ii}
\end{equation}
is the expectation value of the Hamiltonian for a given variational orbital $\varphi_i$, and
\begin{equation}
   \Delta E^\text{Koopmans}_i =
   \begin{cases}
      E^\text{Koopmans}(N) - E^\text{Koopmans}_i(N-1)
       & \text{for occupied orbitals} \\
      E^\text{Koopmans}_i(N+1) - E^\text{Koopmans}(N)
       & \text{for empty orbitals}
   \end{cases}
   \label{eqn:DeltaE_i}
\end{equation}
where $E^\text{Koopmans}_i(N\pm1)$ is the total energy of the system where we add/remove an electron from variational orbital $i$ and allow the rest of the system to relax, with all the other orbitals remaining orthogonal to $\ket{\varphi_i}$.

We use this condition to determine the screening parameters \emph{ab initio}. Specifically, given a starting guess $\{\alpha^0_i\}$ for the screening parameters, an improved guess for the screening parameters can be obtained via
\begin{equation}
   \alpha^{n+1}_i =
   \alpha^n_i \frac{\Delta E^\text{Koopmans}_i - \lambda_{ii}(0, 1)}{\lambda_{ii}(\alpha^n_i, 1) - \lambda_{ii}(0, 1)}; \qquad  \Delta E^\text{Koopmans}_i = E^\text{Koopmans}(N) - E^\text{Koopmans}_i(N - 1)
   \label{eqn:alpha_via_dscf_filled}
\end{equation}
for occupied orbitals and
\begin{equation}
   \alpha^{n+1}_i =
   \alpha^n_i \frac{\Delta E^\text{Koopmans}_i - \lambda_{ii}(0,0)}{\lambda_{ii}(\alpha^n_i,0) - \lambda_{ii}(0,0)}; \qquad \Delta E^\text{Koopmans}_i = E^\text{Koopmans}_i(N+1) - E^\text{Koopmans}(N)
   \label{eqn:alpha_via_dscf_empty}
\end{equation}
for empty orbitals, where
$E^\text{Koopmans}_i(N \pm 1)$ is the total energy of the $N \pm 1$ electron system where we take the $N$-electron system, take this variational orbital $i$ and fill/empty it, and then hold it frozen while the rest of the system is allowed to relax (while remaining orthogonal). These equations yield the screening parameters that satisfy $\lambda_{ii}(\alpha, f) = \Delta E_i^\text{Koopmans}$ if we assume a linear dependence of $\lambda_{ii}$ on $\alpha_i$ and approximate the total energy as a function of $f_i$ to second order. By iterating to self-consistency we lift these approximations and guarantee that $\lambda_{ii}(\alpha, f) = \Delta E_i^\text{Koopmans}$ is satisfied. Typically, only a few iterations are required in order to reach self-consistency, especially if one starts from a physically-motiviated initial guess (such as the static limit of the inverse dielectric function $\varepsilon^{-1}$ in the case of bulk systems). All of these ingredients for calculating $\alpha^{n+1}_i$ are obtained from constrained Koopmans and DFT calculations. Specifically, a $N$-electron Koopmans calculation yields $E^\text{Koopmans}(N)$ and $\lambda_{ii}(\alpha, f)$ (for both $\alpha = \alpha_i^n$ and 0, and $f = 1$ for filled orbitals and $0$ for empty). Meanwhile,  a constrained $N \pm 1$-electron calculation yields $E^\text{Koopmans}_i(N \pm 1)$.

For a periodic system, this method for determining the screening parameters requires a supercell treatment. This is because the $N \pm 1$-electron systems contain a charged defect (because we have filled/emptied a localized orbital) and a supercell is required in order to remove the spurious interactions between periodic images \cite{Nguyen2018, DeGennaro2022}. Section~\ref{sec:dfpt_implementation} will discuss an efficient linear-response reformulation of this problem that avoids a supercell treatment (and can also be used for molecules).

\subsection{Koopmans variants}\label{the-flavour-ki-or-kipz}

As we saw previously in Section~\ref{sec:core_theory}, there is some freedom in how one defines a Koopmans functional. Namely, one must choose values for $\eta_i$, the gradient of the energy as a function of the the occupancy of orbital $i$, for each value of $i$ (modulo the corresponding screening term). In that section, we briefly introduced the Koopmans integer (KI) approach (eq.~\ref{eqn:eta_KI}), but that is just one of several different ways one can define these gradient terms, and it is possible to define several variants.

\subsubsection{KI}\label{ki}

In the KI approach, $\eta_i$ is chosen as the total energy difference of two adjacent electronic configurations with integer occupations as given by the base DFT functional:
\begin{equation}
   \eta_i^{\rm KI} = \left.E^{\rm DFT}\right|_{f_i=1}-\left.E^{\rm DFT}\right|_{f_i=0} = \int_0^{1} \langle \varphi_i \vert \hat{h}^{\rm DFT}(f) \vert \varphi_i \rangle df
\end{equation}
where $\hat{h}^\mathrm{DFT}(f)$ is the DFT Hamiltonian with the occupancy of orbital $i$ constrained to $f$. In this case, the explicit expression for the unscreened KI Koopmans' correction to orbital $i$, which we denote as $\Pi_i^\mathrm{KI}$, becomes
\begin{equation}
   \Pi_i^\mathrm{KI} =
   E_{\rm Hxc} [\rho-\rho_i] -E_{\rm Hxc}[\rho] +f_i \left( E_{\rm Hxc}[\rho-\rho_i+n_i] -E_{\rm Hxc}[\rho-\rho_i] \right)
   \label{eqn:Pi_KI}
\end{equation}
where $\rho_i(\mathbf{r}) = f_i\vert\varphi_i(\mathbf{r})\vert^2$ and $n_i(\mathbf{r}) = \vert\varphi_i(\mathbf{r})\vert^2$. $E_{\rm Hxc}$ denotes the Hartree and exchange-correlation energy corresponding to the underlying base functional.

It can be seen that at integer occupations the KI energy correction vanishes; that is, $\Pi^{\rm KI}_i=0$. In other words, for integer occupations the KI functional preserves the potential energy surface of the base functional! But while the energy correction is vanishing, the potential is non-vanishing --- for example, the KI potential correction to an occupied variational orbital is
\begin{align}
   \frac{\delta \Pi^\mathrm{KI}_{i\sigma}}{\delta \rho_{j\sigma'}(\mathbf{r})} = \left[- E_\mathrm{Hxc}[\rho - n_{i\sigma}] + E_\mathrm{Hxc}[\rho] - \int d\mathbf{r'}\ v^\sigma_\mathrm{Hxc}[\rho](\mathbf{r}') \ n_{i\sigma}(\mathbf{r'})\right] \delta_{ij}\delta_{\sigma\sigma'}
   \label{eqn:dPi_KIdrho}
\end{align}
(here the spin index $\sigma$ has been decoupled from the orbital index). Unlike the energy correction in eq.~\ref{eqn:Pi_KI}, this term is non-zero, which means that the KI correction will affect the spectral properties of the system while leaving the total energy unchanged.

\subsubsection{KIPZ}\label{kipz}

In the KIPZ approach the slope $\eta_i$ is also chosen as the total energy difference of two adjacent electronic configurations with integer occupations, but this time using the Perdew-Zunger (PZ) one-electron-self-interaction corrected (SIC) functional applied to the approximate DFT base functional,
\begin{equation}
   \eta_i^{\rm KIPZ} = \left.E^{\rm PZ}\right|_{f_i=1}-\left.E^{\rm PZ}\right|_{f_i=0} = \int_0^{1} \langle \varphi_i \vert \hat{h}_i^{\rm PZ}(s) \vert \varphi_i \rangle ds,
   \label{eta_kipz}
\end{equation}
In this instance, the explicit expression for the unscreened energy correction corresponding to orbital $i$ (denoted $\Pi_i^{\rm KIPZ}$) becomes
\begin{equation}
   \Pi_i^{\rm KIPZ} = -\int_0^{f_i} \langle \varphi_i \vert \hat{h}^{\rm DFT}(f) \vert \varphi_i \rangle df + f_i \int_0^{1} \langle \varphi_i \vert \hat{h}^{\rm PZ}_i(f) \vert \varphi_i \rangle df,
\end{equation}
where
\begin{equation}
   \hat{h}_i^{\rm PZ}(f) = \hat{h}^{\rm DFT}(f) - \hat{v}^{\rm DFT}_{\rm Hxc}\left[f\vert\varphi_i(\mathbf{r})\vert^2\right]
\end{equation}
is the PZ self-interaction correction applied to the $i^{\rm th}$ variational orbital with constrained occupation $f$, which removes the Hartree-plus-exchange-correlation potential for that orbital. The KIPZ correction can be rewritten as
\begin{equation}
   \Pi^{\rm KIPZ}_i = \Pi^{\rm KI}_i -f_i E_{\rm Hxc} [n_i]
\end{equation}
which makes the physics of this correction clear: it is nothing less than the KI correction with the addition of a (screened) Perdew-Zunger self-interaction correction. This added correction removes one-electron self-interaction and makes the KIPZ functional exact for one-electron systems. In the many-electron case, it provides different (and typically improved) total energies and forces than the base functional\cite{Nguyen2015}, albeit with a screening coefficient for the Perdew-Zunger correction that is inherited from a spectral condition. More details are provided in Supporting Information \ref{sec:kipz_details}.

% In the unscreened case ($\alpha_i = 1$) the KIPZ functional can be thought of as the KI correction applied to the PZ-SIC functional (this can be shown by replacing the base DFT functional and Hamiltonian with its PZ-SIC counterparts). However, in the general case of $\alpha_i \ne 1$ the KIPZ functional form is more general, assigning each PZ self-interaction correction its own screening coefficient. For more details, refer to Ref.~\citenum{Borghi2014}.

\subsubsection{Comparing KI and KIPZ}
\label{sec:pkipz}
The KIPZ correction is more computationally expensive than the KI approach, for the following reasons: we have already mentioned that the KI energy correction vanishes for integer orbital occupations. Furthermore, for occupied orbitals, the KI corrective potential is scalar (i.e.\ it does not have a spatial dependence) and therefore the total energy is invariant with respect to unitary rotations of the variational orbitals. Consequently, once the variational orbitals (and, by extension, the total density) are initialized they remain unchanged during the entire energy minimization procedure. This also implies that the screening parameters of occupied variational orbitals converge instantly (in eq.~\ref{eqn:alpha_via_dscf_filled}, $\Delta E_i$ and $\lambda_{ii}(0, 1)$ are independent of $\alpha^n_i$ and $\lambda_{ii}(\alpha^n_i, 1)$ is linear in $\alpha^n_i$). Contrast this with KIPZ: the KIPZ energy does not match that of the base functional, nor is it invariant with respect to the unitary rotations of occupied orbitals. This means we must directly minimize the energy with respect to the shape of the variational orbitals, greatly increasing the computational cost of these calculations. Furthermore, the KIPZ ground-state density and variational orbitals are a function of the screening parameters, which means that the screening parameters must be calculated self-consistently, further increasing the computational cost.

Despite its additional computational cost, KIPZ has some desirable advantages over KI: for instance, it is one-electron-self-interaction-free. For this reason, we also have introduced the ``perturbative KIPZ'' (pKIPZ) method, where the KIPZ Hamiltonian is applied non-self-consistently to the KI density and variational orbitals, as a way of approximating the KIPZ result at reduced computational cost without significantly compromising the accuracy \cite{Colonna2019}.

It is important to note that the KI functional's invariance with respect to unitary rotations of the occupied variational orbitals introduces an ambiguity in its definition: the variational orbitals are no longer well-defined. This ambiguity is resolved by formally defining the KI functional as the $\gamma \rightarrow 0$ limit of the ``KI$\gamma$PZ" functional, which is the KIPZ functional with the PZ contribution to the correction scaled by a prefactor $\gamma$. This is discussed further in the Supporting Information~\ref{sec:workflows_in_detail_init}.

Finally, we note that the original formulations of Koopmans functionals also introduced the K and the KPZ functionals \cite{Dabo2009,Dabo2010,Borghi2014}. These are similar to the KI and KIPZ functionals, except that the slope $\eta_i$ is evaluated at half-occupation rather than as the total energy difference between integer occupations. These formulations provide almost identical results but more cumbersome than their integer counterparts.

\subsubsection{Total energies and forces with different Koopmans variants}
The design of Koopmans functionals focuses on predicting spectral properties. However, it is worthwhile pausing to consider how accurately these functionals will predict structural properties (namely, total energies and forces). The KI functional, as we have already discussed, yields the same total energy --- and by extension, the same forces --- as its base functional. The KIPZ functional, on the other hand, gives total energies and forces that correspond to its base functional augmented with a screened PZ correction.

There are instances where these two approaches yield significantly different results. For example, in a study of the geometry of adenine, thymine, and uracil, the KIPZ@PBE functional predicted bond lengths with a relative mean absolute error compared to experiment of $0.65\%$, which was slightly better than PBE0 ($0.76\%$) and PZ@PBE ($0.83\%$), and was markedly better than PBE ($1.63\%$) --- and, by extension, KI@PBE \cite{Nguyen2016}. That same study showed that the KIPZ@PBE functional captured the tilt of the amino groups of nucleobases with respect to their aromatic rings, whereas PBE wrongly predicts a near-planar structure. However, the addition of a PZ correction does not necessarily improve structural properties across the board. Ref.~\citenum{Borghi2014} compared structural properties for the reference G2-1 set of molecules, and found that KIPZ@PBE predicted bond angles less accurately (with a mean relative error of 2.2\% for KIPZ@PBE compared to 1.4\% for PBE) despite predicting bond lengths slightly better (1.5\% for KIPZ@PBE compared to 2.3\% for PBE).

We stress that these considerations regarding structural properties are somewhat orthogonal to the Koopmans functional formalism. One should not use the KI functional to calculate structural properties alone (because the ODD formalism comes at increased computational cost but provides no change in the structural properties). If desired, improved geometrical properties and accurate spectral properties can be simultaneously obtained by combining the KI correction with a more advanced base functional that predicts structural properties more reliably.

\subsection{Important caveats}
\label{sec:other_considerations}
Before concluding this section, there are a few further important points that must be made.

\subsubsection{Restriction to systems with a non-zero band gap}
\label{sec:nonzero_bandgap}
First, the Koopmans formulation is only well-defined for systems with a non-zero band gap. This is because the Koopmans correction (eq.~\ref{eqn:lambda_ii}) is defined in terms of the diagonal elements of the occupation matrix. A band gap (however small) means that the occupancy matrix is block-diagonal, and can always be chosen to be the identity for the occupied manifold and zero for the unoccupied manifold. In the absence of a band gap, the occupancy matrix is not block-diagonal and a well-defined Koopmans functional would require some (currently unknown) corrections for the off-diagonal components. While it would be desirable to derive an off-diagonal correction and to lift this restriction, the current theory remains powerful --- after all, it is in insulating and semi-conducting systems where DFT exhibits one of its most striking failures in the underestimation of the band gap.

However, we note that we often we rely on semi-local DFT as the base functional to define or initialize the variational orbitals. If the base functional also predicts a non-zero band gap, then the valence and conduction manifold can be disentangled \cite{Qiao2023a}, the occupancy matrix will be block-diagonal, and the Koopmans correction can immediately be applied. However, if the base functional wrongly predicts a metallic state, then the valence and conduction manifolds are not so easily disentangled. In these cases, one might be able to first employ other base functionals to open a gap (such as DFT\,+\,\emph{U}) or deploy novel projectability disentanglement methods to separate the valence and conduction manifolds \cite{Qiao2023}.

The occupancies of variational orbitals $f_i$ have been a central quantity in constructing the Koopmans formalism. This restriction to systems with a band gap means that these variational orbital occupancies will always be either 0 or 1, and consequently some terms in the formalism vanish (for example, the KI correction to the energy; eq.~\ref{eqn:Pi_KI}) but others do not (for example, the KI correction to the potential; eq.~\ref{eqn:dPi_KIdrho}).

\subsubsection{Empty state localization in the bulk limit}
While minimizing the Koopmans energy functional for bulk systems leads to well-localized occupied orbitals, the same process does not lead to well-localized empty orbitals. This is because (a) low-lying conduction bands are often entangled with highly-delocalized nearly-free-electron bands, and (b) the Koopmans correction to empty states contains a leading Hartree term that incentivizes delocalization (see Ref.~\citenum{Borghi2014}). However, the Koopmans correction ought to be applied to localized orbitals, and vanishes in the limit of infinitely delocalized states (as discussed in Section~\ref{sec:odd}). In light of this, we typically apply the Koopmans correction non-self-consistently on a maximally localized Wannier function representation of the empty manifold. This approach is heuristic but effective, as demonstrated by previous works \cite{Nguyen2018,DeGennaro2022}.

\subsubsection{Symmetries}
Because a Koopmans potential $v^\mathrm{Koopmans}[\rho,\rho_i]$ is constructed via a variational orbital density, these potentials can break the translational symmetry of periodic systems. However,
the variational orbitals crucially possess the translational properties of Wannier functions; that is, for each variational orbital $\varphi_{\mathbf{R}}$ there exists a periodic replica $\varphi_{\mathbf{R+R'}}$ such that
\begin{equation}
   \varphi_\mathbf{R}(\mathbf{r} - \mathbf{R'})
   =
   \varphi_{\mathbf{R} + \mathbf{R'}}(\mathbf{r})
\end{equation}
where $\mathbf{R}$ and $\mathbf{R}'$ can be any pair of Bravais lattice vectors. Thanks to this property, the collective potential $\sum_i v^\mathrm{Koopmans}[\rho, \rho_i] |\varphi_i \rangle \langle \varphi_i|$ inherits the translational symmetry of the overall system and thus it remains possible to describe the system's electronic structure with a band-structure picture. For more details, refer to Ref.~\citenum{DeGennaro2022}.

More generally, the orbital-density dependence of Koopmans functionals might unphysically break the crystal point group symmetry. This is a common feature of non-rotationally invariant methods that are based on localized orbitals \cite{Stengel2008,Lehtola2016}. Here, the symmetry of the localized representation plays an important role, especially in small systems and in the atomic limit. Possible solutions to this issue have been recently suggested\cite{Su2020}, and this point is worthy of further investigation.

\section{Algorithms and implementation}
\label{sec:implementation}
% The Koopmans formalism that underpins all of these calculations is inherently more complex than a semi-local DFT algorithm. In a (semi-)local DFT calculation, all that one needs to do is determine the ground-state density. For Koopmans calculations one must also obtain the minimizing variational orbitals as well as the screening parameters. This requires custom implementation within electronic structure codes.

The formulation of Koopmans functionals, as outlined in the previous section, is inherently more complex than a ``standard" semi-local DFT calculation, and requires non-standard algorithms and bespoke implementation within electronic-structure codes. This section describes these algorithms and describes how Koopmans functionals have been implemented in \texttt{\textsc{Quantum ESPRESSO}} and the open-source package \texttt{koopmans}.

\subsection{Orbital optimization}
\label{sec:dscf_implementation}
In order to work with Koopmans functionals, we must be able to minimize an orbital-density-dependent functional. In other words, we must optimize a set of orbital densities $\{\rho_i\}$ such that the Koopmans energy functional (eq.~\ref{eqn:Koopmans_functional_form}) is minimized. This orbital optimization is performed separately for the occupied and then the empty manifold using an optimization algorithm similar to that employed in the ensemble DFT approach\cite{Marzari1997}: the orbital densities are parameterized via a set of wavefunctions $\phi_i$ and a unitary rotation matrix $U$, such that $\rho_i = |(U \phi)_i|^2$, and then the energy is then minimized via the nested loop:
\begin{equation}
   E = \min_{\{\phi_i\}} \left(\min_{U} E^\mathrm{Koopmans}[\{|(U\phi)_i|^2\}] \right)
\end{equation}
where in the inner loop the unitary rotation matrix $U$ is optimized (which leaves the total density unchanged), and in the outer loop the wave functions are optimized. Both steps are performed using the conjugate-gradient algorithm. The optimization is performed separately for the occupied and empty manifolds to ensure that the occupation matrix remains block-diagonal (as discussed in Section~\ref{sec:nonzero_bandgap}).

One important ingredient in ODD energy minimization is the use of complex orbitals. Because the ODD energy is not invariant with respect to unitary rotations of the variational orbitals, it can no longer be assumed (as in the case for DFT) that the variational orbitals are real, and thus the aforementioned wave functions $\phi_i$ must be complex in order to find the true minimum of the ODD functional.\cite{Borghi2014,Klupfel2011,Hofmann2012,Lehtola2014,Lehtola2016}

In addition to the generic orbital minimization procedure, we must also perform constrained minimization calculations (as required by the finite-difference method for calculating screening parameters; Section~\ref{sec:calculating_alphas}). Here, the total ODD energy is minimized while removing/adding one electron to a particular variational orbital. (This gives us $E_i(N\pm1)$ from eq.~\ref{eqn:DeltaE_i}). This orbital must be frozen during the minimization, otherwise it would morph into the valence band maximum/conduction band minimum, and one must also impose the standard orthogonality condition with all other orbitals belonging to the same spin channel. Image correction methods such as Martina-Tuckerman or Gygi-Baldereschi \cite{Martnya1999a,Gygi1986} must be used to avoid spurious interaction between charged periodic images. For periodic systems this also means that these calculations must be performed in a supercell. These charged defect calculations also require special care in low-dimensional materials \cite{Komsa2014}. Further details regarding the orbital minimization procedure are presented in Ref.~\citenum{Borghi2015}.

\subsection{The \texttt{kcp.x} code}
These orbital minimization algorithms are implemented in the code \texttt{kcp.x}. In other words, \texttt{kcp.x} can be used to obtain the ground-state energy and the minimizing set of variational orbitals of an arbitrary system for a given orbital-density-dependent functional (PZ, KI, or KIPZ).

\texttt{kcp.x} can be used to calculate screening parameters via the finite-difference approach, and is applicable to both periodic and aperiodic systems. By design, it does not use $\mathbf{k}$-point sampling for periodic systems, because the finite-difference approach mandates the use of a supercell (as discussed above), rendering $\mathbf{k}$-point sampling of the Brillouin zone superfluous. Instead, the dimensions of the supercell can be used to effectively sample $\mathbf{k}$-space for bulk systems, and the band structure for the equivalent primitive cell can be reconstructed at the end of the calculation using an unfolding procedure \cite{DeGennaro2022}. Despite the absence of $\mathbf{k}$-space sampling (which is embarrassingly parallel), the \texttt{kcp.x} code still uses MPI parallelism: it is parallelized over the plane wave basis. This allows for the distribution of linear algebra operations and Fourier transforms across processors.

Because Koopmans functionals are a correction applied on top of a local or semilocal functional, and these functionals are computationally inexpensive compared to their ODD counterparts, before commencing orbital minimization with \texttt{kcp.x} it is efficient to initialize the variational orbitals as Kohn-Sham orbitals or maximally localized Wannier functions \cite{Marzari2012}. To support the use of Wannier functions for periodic systems, we have implemented an interface that takes set of $\mathbf{k}$-indexed Wannier functions from a \texttt{Wannier90} calculation and maps it to an enlarged set of $\Gamma$-only Wannier functions defined on the corresponding supercell.
% The image correction methods currently implemented in \texttt{kcp.x} assume an isotropic dielectric $\varepsilon$. In the near future these will be extended to support a fully tensorial $\boldsymbol{\varepsilon}$, which is critical for --- among other things --- calculations on two-dimensional materials.
Given that \texttt{kcp.x} implements the full minimization of the ODD functional, in principle one could use the output of \texttt{kcp.x} to perform geometry optimizations, calculate phonons via the frozen-phonon method, calculate electron-phonon coupling, model excitons, and so on.

% Discuss v4.1
For historical reasons, \texttt{kcp.x} is implemented on top of \texttt{cp.x}, the code within \texttt{Quantum ESPRESSO} usually responsible for performing Car-Parrinello molecular dynamics (hence the name ``\texttt{kcp.x}"), which already contained algorithms similar to the direct functional minimization required by Koopmans functionals. It is important to note that \texttt{kcp.x} is not meant to perform molecular dynamics like \texttt{cp.x}. The implementation is built on top of version 4.1 of \texttt{Quantum ESPRESSO}. The modifications made to implement Koopmans functionals are (a) extensive and (b) of no relevance to the standard functioning of the \texttt{cp.x} code, so these modifications have not yet been incorporated within the official \texttt{Quantum ESPRESSO} repository, nor was the private version of the code kept aligned with subsequent \texttt{Quantum ESPRESSO} releases. Fast-forward to today, and \texttt{kcp.x} has effectively become a standalone code.

\subsection{Screening parameters via linear response calculations in reciprocal space}
\label{sec:dfpt_implementation}

While the finite-difference approach of \texttt{kcp.x} can provide us with all of the ingredients to calculate the screening parameters, it is somewhat cumbersome, since one must perform several constrained DFT and Koopmans calculations, and for periodic systems these must be performed in a supercell. An alternative to this approach is to compute the screening coefficients via density-functional perturbation theory (DFPT) \cite{Baroni2001}.

In this approach, one first approximates the energy as a quadratic function of the occupation number (which is typically a very good approximation), and the expression for the screening coefficients reduces to
\begin{equation}
   \alpha_i = \frac{d^2E_{\rm DFT}/df_i^2}{\partial^2 E_{\rm DFT}/\partial f_i^2} = \frac{\langle n_i \vert \epsilon^{-1} f_{\rm Hxc} \vert n_i \rangle}{\langle n_i \vert f_{\rm Hxc} \vert n_i \rangle},
   \label{eqn:alpha_via_dfpt}
\end{equation}
where $\frac{d}{df_i}$ ($\frac{\partial}{\partial f_i}$) represents variations that do (do not) account for orbital relaxation, $\epsilon(\mathbf{r},\mathbf{r}')$ is the microscopic dielectric function of the material, $f_{\rm Hxc}(\mathbf{r},\mathbf{r}') = \delta^2 E_{Hxc}/ \delta \rho(\mathbf{r})\delta \rho(\mathbf{r}')$ is the Hartree-plus-exchange-and-correlation kernel, and $n_i(\mathbf{r})=\vert\varphi_i(\mathbf{r})\vert^2$ is the orbital density at integer occupation\cite{Colonna2018}. This can be evaluated by considering the density response $\Delta^{i} n(\mathbf{r})$ induced in the system by the perturbing potential $v^{i}_{\rm pert}(\mathbf{r}) = \int d\mathbf{r}' f_{\rm Hxc}(\mathbf{r}, \mathbf{r}') n_{i}(\mathbf{r}')$. This perturbation is the Hartree-plus-exchange-and-correlation potential generated when adding/removing an infinitesimal fraction of an electron to/from orbital $i$. One determines $\Delta^i n$ self-consistently via DFPT\cite{Colonna2022}, and then the screening parameters are given by
\begin{equation}
   \alpha_{i} = 1 + \frac{\langle v^{i}_{\rm pert} \vert \Delta^{i} n \rangle}{\langle n_{i} \vert v^{i}_{\rm pert} \rangle}.
\end{equation}
Evaluating the screening coefficients within this linear-response approach only requires quantities available from a $N$-electron calculation, which means that in the case of periodic solids there is no need for a supercell. Instead, we can reduce the cost of these calculations by taking advantage of the translational symmetry of the system \cite{DeGennaro2022} and recasting the supercell problem in a basis of Wannier functions. These Wannier functions take the form $w^{\mathbf{R}i}(\mathbf{r})$, where the orbital label explicitly denotes the lattice vector $\mathbf{R}$ of the home cell inside the supercell. In this basis, the DFPT expression for the screening coefficients (eq.~\ref{eqn:alpha_via_dfpt}) can be decomposed into a set of independent problems (monochromatic perturbations), one for each $\mathbf{q}$ point sampling the Brillouin zone of the primitive cell.\cite{Colonna2022} The now $\mathbf{q}$-dependent charge density variation $\Delta n^{\mathbf{0}i}_\mathbf{q}(\mathbf{r})$ induced by the perturbing potential $v^{\mathbf{0}i}_{\rm pert,\mathbf{q}}$ is obtained self-consistently via DFPT (eqs.~15-17 of Ref.~\citenum{Colonna2022}), and then the screening coefficients are obtained by summing over $\mathbf{q}$:
\begin{equation}
   \alpha_{\mathbf{0}i} =  1 + \frac{\sum_{\mathbf{q}} \langle v^{\mathbf{0}i}_{\rm pert,\mathbf{q}} \vert \Delta^{\mathbf{0}i}_{\mathbf{q}}n \rangle} {\sum_{\mathbf{q}} \langle n^{\mathbf{0}i}_{\mathbf{q}} \vert v^{\mathbf{0}i}_{\rm pert,\mathbf{q}} \rangle}.
   \label{eqn:alpha_via_dfpt_kspace}
\end{equation}
The KI Hamiltonian at a particular $\mathbf{k}$ point is then given to second order by
\begin{equation}
   H_{i j}^{\mathrm{KI}(2)}(\mathbf{k})=H_{i j}^{\mathrm{DFT}}(\mathbf{k})+\alpha_{\mathbf{0} j} \Delta H_{i j}^{\mathrm{KI}(2)}(\mathbf{k})
   \label{eqn:ki_hamiltonian_kcw}
\end{equation}
where the second-order KI contribution to the Hamiltonian is
\begin{equation}
   \Delta H_{v v^{\prime}}^{\mathrm{KI}(2)}(\mathbf{k})
   =-\frac{1}{2} \left\langle n_{\mathbf{q}}^{\mathbf{0}v} \middle| v_{\rm pert, \mathbf{q}}^{\mathbf{0}v}\right\rangle \delta_{v v^{\prime}}
\end{equation}
for valence bands and
\begin{equation}
   \Delta H_{c c^{\prime}}^{\mathrm{KI}(2)}(\mathbf{k})=-\frac{1}{2}
   \left\langle n_{\mathbf{q}}^{\mathbf{0}c} \middle| v_{\rm pert,\mathbf{q}}^{\mathbf{0}c}\right\rangle
   \delta_{c c^{\prime}}+\frac{1}{N_{\mathbf{q}}} \sum_{\mathbf{q}}\left\langle v_{\mathrm{pert}, \mathbf{q}}^{\mathbf{0} c^{\prime}} \middle| n_{\mathbf{k}, \mathbf{k}+\mathbf{q}}^{c c^{\prime}}\right\rangle
   \label{eqn:delta_h_ki_dfpt}
\end{equation}
for conduction bands, where $n^{cc'}_{\mathbf{k},\mathbf{k}+\mathbf{q}}(\mathbf{r})=(w^{c}_\mathbf{k}(\mathbf{r}))^*w^{c'}_{\mathbf{k}+\mathbf{q}}(\mathbf{r})$; $w_\mathbf{k}^c(r)$ the periodic part of the electronic state in the Wannier gauge. As expected, the KI contribution to the valence bands is $\mathbf{k}$-independent. The total Hamiltonian is then diagonalized in order to obtain the canonical eigenstates and energies. Given the fact that the Hamiltonian is written in a basis of Wannier functions, it is also possible to employ standard interpolation techniques to obtain the KI eigenvalues at any arbitrary $\mathbf{k}$-point\cite{Marzari2012}.

However, the DFPT approach does come with some limitations. The principal limitation is that the energy is approximated to second order in the perturbing potential. In most cases this is very accurate, correctly capturing the quadratic Hartree contribution and only missing the non-quadratic, higher-order exchange-correlation contributions.

\subsection{The \texttt{kcw.x} code}
The calculation of screening parameters via DFPT and the subsequent construction of the Koopmans Hamiltonian and band structure, as described above, has been implemented in the code \texttt{kcw.x}. Because all of these calculations are performed in a basis of Wannier functions, this code obtains Wannier functions via an interface with \texttt{Wannier90}. (The ``\texttt{w}'' in \texttt{kcw} stands for ``Wannier''.) Because all of these equations are formulated in terms of a primitive cell with $\mathbf{k}$-point sampling, \texttt{kcw.x} uses MPI to parallelize over $\mathbf{k}$-points. It also parallelizes over plane-wave orbitals (as already introduced in the context of \texttt{kcp.x}).

While much of the above applies to periodic systems, \texttt{kcw.x} can still be used to perform calculations on aperiodic systems. The Wannier function basis still remains valid, but we no longer have multiple $\mathbf{k}$-points.

\texttt{kcw.x} is part of the official \texttt{Quantum ESPRESSO} distribution (from version 7.1 onward).

\subsection{Comparing \texttt{kcp.x} and \texttt{kcw.x}}
\texttt{kcp.x} and \texttt{kcw.x} implement different Koopmans strategies and, as such, they have different use-cases, largely defined by their computational scaling. The two codes scale differently largely due to the fact that \texttt{kcw.x} operates in a primitive cell while \texttt{kcp.x} operates in a supercell. Calculating one screening parameter using \texttt{kcp.x} requires multiple SCF calculations, each of which takes a computational time $T^{\rm SC}$ that roughly scales as $\mathcal{O}\left(\left({N_{\rm orb}^{\rm SC}}\right)^3\right)$, where $N_{\rm orb}^{\rm SC}$ is the number of orbitals in the supercell. Meanwhile, calculating one screening parameter using the \texttt{kcw.x} DFPT approach scales as $T^{\rm PC} \propto N_{\mathbf{q}} N_{\mathbf{k}} {N_{\rm orb}^{\rm PC}}^3$. This is the typical computational time for the SCF cycle $N_{\mathbf{k}} {N_{\rm orb}^{\rm PC}}^3$ times the number of independent monochromatic perturbations $N_{\mathbf{q}}$.  Using the relation $N_{\rm orb}^{\rm SC}=N_{\mathbf{k}}N_{\rm orb}^{\rm PC}$ and the fact that $N_{\mathbf{q}} \lesssim N_{\mathbf{k}}$, the ratio between the supercell and primitive computational times is roughly proportional to $N_{\mathbf{q}}$. Thus, as the supercell size (or equivalently the number of $\mathbf{q}$-points in the primitive cell) increases, the \texttt{kcw.x} DFPT approach becomes more and more computationally efficient.\cite{Colonna2022} For aperiodic systems, $N_\mathbf{q}=1$ and the two approaches scale similarly, but with different prefactors.

Note that these scaling relations pertain to the calculation of a single screening parameter, whereas a full Koopmans workflow requires the calculation of one screening parameter per unique variational orbital in the system. Here, the word ``unique" is very important; orbitals that are related by symmetry will share the same screening parameter and therefore the screening does not need to be recalculated for each orbital. This means that in the worst-case scenario, where none of the variational orbitals are related by symmetry, the overall scaling of the workflow has an additional $N_{\rm orb}$ prefactor, but for many systems (and for periodic systems in particular) the number of unique variational orbitals in the system can be many times smaller than the total number of orbitals. Furthermore, (a) the calculation of screening parameters for separate orbitals is embarrassingly parallelizable, and (b) it is possible to predict the screening parameters via machine learning, avoiding the need to repetitively calculate screening parameters altogether \cite{Schubert2022}.

The superior scaling of \texttt{kcw.x} comes at a cost, as it makes two approximations that \texttt{kcp.x} does not: the DFPT approach expands the total energy only to second order when computing screening parameters (see Section~\ref{sec:dfpt_implementation}), and it does not optimize the variational orbitals. These are instead defined via Wannier functions, which often closely resemble the minimizing orbitals of the Koopmans energy functional. This also means that \texttt{kcw.x} only implements the KI functional. Without orbital minimization one cannot perform KIPZ calculations, and pKIPZ would require the PZ kernel (i.e. the second derivative of the PZ energy with respect to the density), and this is not implemented in common electronic-structure codes.

\subsection{Workflow management}
Running a Koopmans calculation with either \texttt{kcp.x} or \texttt{kcw.x} requires a few additional steps compared to a standard semi-local DFT calculation. In this section, we will focus on the workflows that one needs to perform in order to complete a Koopmans functional calculation, and how these are publicly disseminated in open-source form.

Typically, these workflows can be divided into three steps:

\begin{enumerate}
   \item
         an initialization step, where the variational orbitals are initialized
   \item
         the calculation of screening parameters
   \item
         a final calculation using the final screening parameters
\end{enumerate}
Depending on the method used for calculating screening parameters (that is, either finite differences with \texttt{kcp.x} or DFPT with \texttt{kcw.x}), the resulting workflows look very different. Differences also emerge between calculations on molecules and solids. For the latter (and for large molecular systems), we have already seen that maximally localized Wannier functions are typically used as the variational orbitals (for KI) or as a starting guess for the variational orbitals (for KIPZ). This necessitates an additional Wannierization procedure \cite{Marzari2012} and an interface between \texttt{Wannier90} and \texttt{kcp.x}/\texttt{kcw.x}. Meanwhile, for calculating the screening parameters via finite differences, we must perform a combination of different constrained orbital minimizations. In all cases, the workflows typically comprise of several if not dozens of calculations, often involving different electronic structure codes that must handshake with one another. This can greatly benefit from automation.

\subsection{The \texttt{koopmans} package} These workflows are all implemented within the \texttt{koopmans} package. Users exclusively interact with \texttt{koopmans}, rather than the electronic structure codes directly (which can include, in addition to \texttt{kcp.x} and \texttt{kcw.x}, pre-existing codes such as \texttt{pw.x}, \texttt{pw2wannier90.x}, and \texttt{wannier90.x} \cite{Giannozzi2009,Giannozzi2017,Pizzi2020}).

Typically, a user provides \texttt{koopmans} with a single input \texttt{JSON} file (some examples are provided in Supporting Information~\ref{sec:io_files}). Based on the settings provided in this input file, \texttt{koopmans} proceeds through the requested workflow. Whenever an electronic structure calculation needs to be performed, it generates the corresponding input file, calls the relevant code, waits for it to complete, and then parses the output file. Between successive calculations, it computes intermediate variables, moves and modifies files, \emph{etc.}. In other words, the workflow runner takes care of the banal aspects of performing a Koopmans calculation, allowing users to concern themselves with scientific matters (e.g. ``what functional do I want to use?") rather than getting bogged down in practical details (e.g. ``are the Wannier function files in the correct format for the next calculation to be able to read?")

The \texttt{koopmans} package is shipped with versions of \texttt{Quantum ESPRESSO} that contain \texttt{kcp.x} and \texttt{kcw.x}, meaning that it contains everything that is required to perform Koopmans functional calculations from start to finish.

Further details on the \texttt{koopmans} package can be found in Supporting Information~\ref{sec:koopmans_code_details}. A step-by-step explanation of the workflows themselves can be found in Supporting Information~\ref{sec:workflows_in_detail}.

% This package comprises of three distinct sets of codes. These are two independent implementations of Koopmans functionals in \texttt{\textsc{Quantum ESPRESSO}} \cite{Giannozzi2009,Giannozzi2017}, and are a python package that automates the Koopmans workflows.

\section{Example calculations}
\label{sec:examples}
This Koopmans functional formalism has already proven to be very powerful. In Ref.~\citenum{Colonna2019}, Koopmans functionals were found to predict the ionization potentials of a set of 100 small molecules with comparable/superior accuracy to state-of-the-art GW approaches. Importantly, Koopmans functionals do not only correct the ionization potential (i.e.\ the charged excitation where the most weakly bound electron is removed) but \emph{any} single-particle charged excitation. This was shown for a large set of molecules relevant for photovoltaic applications \cite{Nguyen2015}, with Koopmans functionals yielding ultraviolet photoemission spectra that agree quantitatively with experiment. One can see similar accuracy in the prediction of band gaps and band structures of periodic systems\cite{Nguyen2018,DeGennaro2022,Colonna2022}; in a study of prototypical semiconductors and insulators, Koopmans functionals were found to yield band gaps with a mean absolute error of 0.22\,eV, compared to 0.18\,eV when using self-consistent GW with vertex corrections \cite{Nguyen2018}. Importantly, alignment between the valence band edge and the vacuum level was also very good: across six semiconductors the mean absolute error was 0.19\,eV, compared to 0.39\,eV for G$_0$W$_0$ and 0.49\,eV for self-consistent GW with vertex corrections. Finally, Koopmans functionals can accurately describe the spectral properties of liquids, with the KIPZ functional predicting the electronic density of states of liquid water with comparable accuracy to self-consistent GW with vertex corrections \cite{deAlmeida2021}.

However, all of these calculations were performed by individuals with expert knowledge of Koopmans functionals and with specific expertise on the codes that implement them. This final section demonstrates the capabilities of the \texttt{koopmans} package by way of several examples. All of the following calculations are possible using a very minimalist input file (see Supporting Information~\ref{sec:io_files}). Note that the following calculations use slightly underconverged parameters (specifically, the energy cutoff, cell size, and/or the size of the $k$-point grid). Our focus here is to provide example calculations that can be reproduced easily by readers, rather than providing high-quality reference results.

\subsection{The ionization potential and electron affinity of ozone}
\label{sec:ozone}
First, we present the calculation of the ionization potential and electron affinity of ozone using \texttt{koopmans}.

This calculation is run with the simple command \texttt{koopmans ozone.json}; the input and output files for which can be found in Supporting Information~\ref{sec:ozone_io_files}. In short, this command prompts the full sequence of \texttt{Quantum ESPRESSO} calculations necessary to initialize the density and variational orbitals, calculate the screening parameters, and run a final KI calculation. The \texttt{Quantum ESPRESSO} input and output files for these calculations are all stored in various subdirectories of the current working directory. In principle one can then simply parse the quantities of interest from the output files (but there are easier ways, as explained in Supporting Information~\ref{sec:scriptability}). Refer to Supporting Information~\ref{sec:dscf_workflow_in_detail} for a detailed step-by-step description of this workflow.

\begin{table}[t]
   \centering
   \begin{tabular}{l d{5.5} d{7.7} l}
      \hline
                                           & \multicolumn{1}{c}{IP} & \multicolumn{1}{c}{EA}                                                                    \\
      \hline
      PBE                                  & 7.95                   & 6.17                   & This work                                                        \\
      % KI@[PBE,PZ]      & 13.16   & 1.90 \\
      % pKIPZ@[PBE,PZ]   & 12.68   & 0.42?!
      % KIPZ@[PBE,PZ]    & 12.18   & -9.53?!
      G\textsubscript{0}W\textsubscript{0} & 11.80 \pm 0.25         & 2.34 \pm 0.25          & Ref.~\citenum{vanSetten2015}                                     \\
      scGW\textsubscript{0}@PBE            & 12.57                  &                        & Ref.~\citenum{Caruso2016}                                        \\
      scGW\textsubscript{0}@HF             & 13.16                  &                        & Ref.~\citenum{Caruso2016}                                        \\
      scGW                                 & 12.54                  &                        & Ref.~\citenum{Caruso2016}                                        \\
      qsGW                                 & 13.21                  &                        & Ref.~\citenum{Caruso2016}                                        \\
      CCSD(T)                              & 12.55                  &                        & Ref.~\citenum{Krause2015}                                        \\
      KI@[PBE,KS]                          & 12.52                  & 1.82                   & This work                                                        \\
      KI@[PBE,KS]                          & 12.91                  &                        & Ref.~\citenum{Colonna2019}                                       \\
      experiment                           & 12.73                  & 2.10                   & Refs.~\citenum{NIST_webbook,Katsumata1984,Novich1979,Arnold1994} \\
      \hline
   \end{tabular}
   \caption{The vertical ionization potential (IP) and electron affinity (EA) of ozone, as calculated using functional, perturbative, and quantum chemistry methods, as well as experiment. KI@[PBE,KS] denotes the KI Koopmans correction on top of the PBE base functional, with Kohn-Sham orbitals defining the variational orbitals, showing excellent performance compared to state-of-the-art methods. The uncertainties in the G\textsubscript{0}W\textsubscript{0} correspond to the standard deviation of values reported in Ref.~\citenum{vanSetten2015}, which presents calculations using a range of codes and basis sets.}
   \label{tab:ozone_ip_and_ea}
\end{table}

The ionization potential (IP) and electron affinity (EA) of ozone, as given by this calculation, are listed in Table~\ref{tab:ozone_ip_and_ea}, showing the excellent performance of the KI functional compared to state-of-the-art methods.

\subsection{The band structure of silicon}
\texttt{koopmans} can also perform calculations on bulk systems. Here one typically performs a Wannierization procedure in order to generate maximally localized Wannier functions to use as variational orbitals. Running this calculation gives rise to a similar output to the previous case, with the notable exception that the initialization procedure now involves Wannierization (see Supporting Information~\ref{sec:silicon_io_files}).

\begin{figure}[t]
   \centering
   \includegraphics[width=\columnwidth]{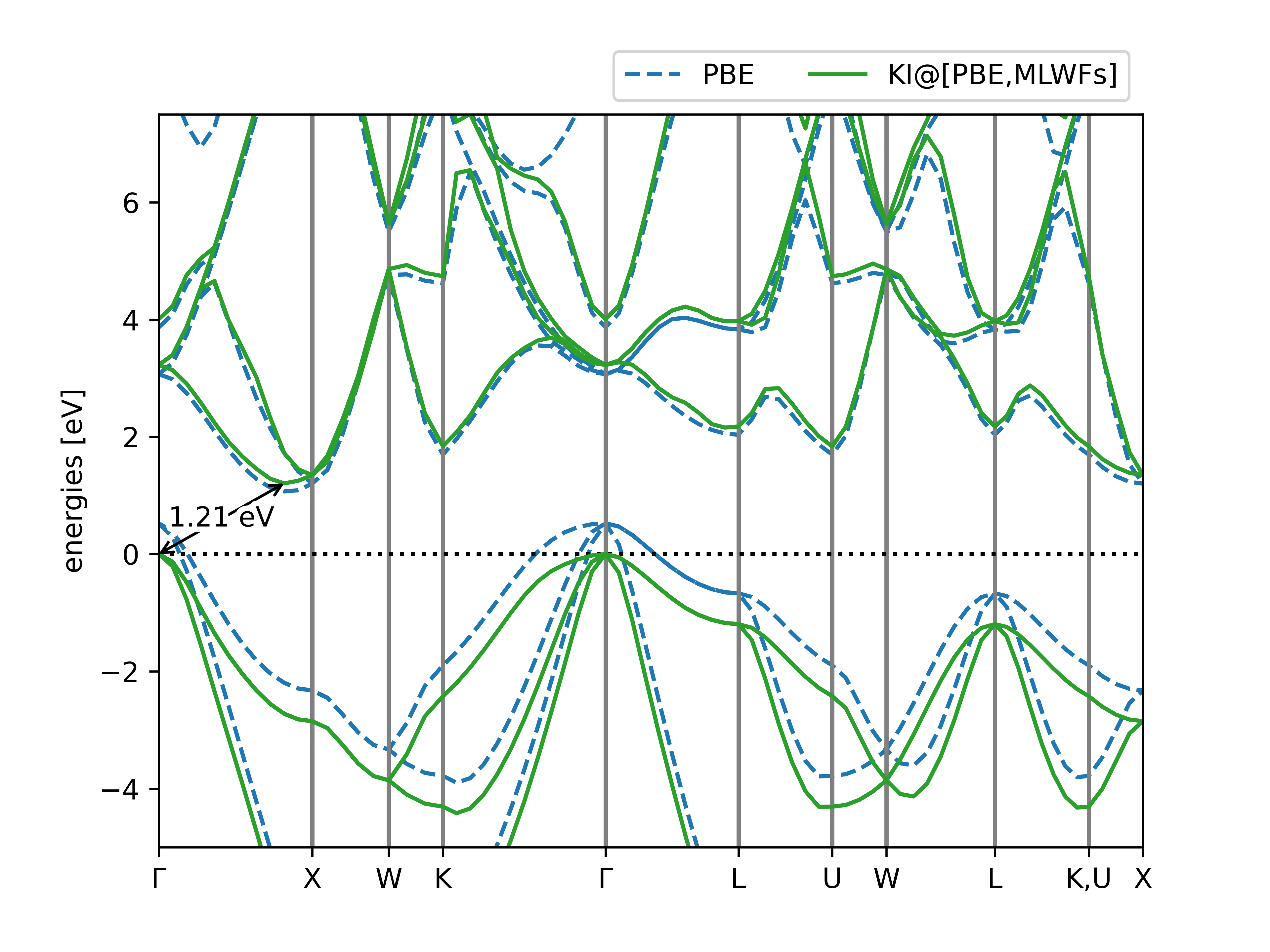}
   \caption{Band structure of bulk silicon, calculated using the KI functional with the PBE base functional, MLWFs as variational orbitals, and screening parameters calculated via finite differences. The PBE band structure is also plotted for comparison.}
   \label{fig:si_band_struct}
\end{figure}

The band structure that one obtains from this calculation is shown in Figure~\ref{fig:si_band_struct}, the band gap is displayed in Table~\ref{tab:si_band_struct}, alongside energy differences between particular symmetry points in the band structure.

The experimental band gap is reproduced with accuracy comparable to self-consistent GW with vertex corrections, and the energy differences between symmetry points are reproduced with comparable accuracy to G\textsubscript{0}W\textsubscript{0} (the only perturbative method for which these data were available). Note that the PBE and KI valence-to-valence energy differences match. This occurs because the occupied manifold is comprised of four identical Wannier functions, and thus the KI correction to these bands amounts to a rigid shift. Contrast this with the valence-to-conduction energy differences, which are markedly better for the KI functional.

\begin{table}[t]
   \centering
   %%diff%%\color{blue}
   \begin{tabular}{r@{ $\rightarrow$ } l *{6}{d{2.2}} d{2.2} @{$\pm$} d{1.2}}
      \hline
      \hline
      \multicolumn{2}{c}{ }
                                & \multicolumn{1}{c}{\multirow{2}{*}{PBE\textsuperscript{\emph{a}}}}
                                & \multicolumn{1}{c}{\multirow{2}{*}{G\textsubscript{0}W\textsubscript{0}\textsuperscript{\emph{b}}}}
                                & \multicolumn{1}{c}{\multirow{2}{*}{scG$\tilde{\mathrm{W}}$\textsuperscript{\emph{c}}}}
                                & \multicolumn{2}{c}{KI@[PBE,MLWFs]}
                                & \multicolumn{1}{c}{\multirow{2}{*}{KIPZ@PBE\textsuperscript{\emph{d}}}}
                                & \multicolumn{2}{c}{\multirow{2}{*}{exp\textsuperscript{\emph{e}}}}                                                                                                                                                                   \\
      \multicolumn{2}{c}{ }     &                                                                           &       &       & \multicolumn{1}{c}{this work} & \multicolumn{1}{c}{Ref.~\citenum{DeGennaro2022}}                               \\
      \hline
      \multicolumn{2}{c}{$E_g$} &
      0.49 &  1.06 & 1.14 &  1.16 &  1.12 &  1.15 & \multicolumn{2}{c}{1.17}\\
      $\Gamma_{1v}$ & $\Gamma_{25'v}$ & 11.97 & 12.04 &      & 11.97 & 11.96 & 12.09 & 12.5 &  0.6\\
      $X_{1v}$ & $\Gamma_{25'v}$ &  7.82 &       &      &  7.82 &       &       & \multicolumn{2}{c}{7.75}\\
      $X_{4v}$ & $\Gamma_{25'v}$ &  2.85 &  2.99 &      &  2.85 &  2.84 &  2.86 & \multicolumn{2}{c}{2.90}\\
      $L_{2'v}$ & $\Gamma_{25'v}$ &  9.63 &  9.79 &      &  9.63 &  9.63 &  9.74 &  9.3 &  0.4\\
      $L_{1v}$ & $\Gamma_{25'v}$ &  6.98 &  7.18 &      &  6.98 &  6.96 &  7.04 &  6.8 &  0.2\\
      $L_{3'v}$ & $\Gamma_{25'v}$ &  1.19 &  1.27 &      &  1.19 &       &       &  1.2 &  0.2\\
      $\Gamma_{25'v}$ &  $\Gamma_{15c}$ &  2.48 &  3.29 &      &  3.17 &  3.18 &  3.20 & 3.35 & 0.01\\
      $\Gamma_{25'v}$ &  $\Gamma_{2'c}$ &  3.28 &  4.02 &      &  3.95 &  3.94 &  3.95 & 4.15 & 0.05\\
      $\Gamma_{25'v}$ &        $X_{1c}$ &  0.62 &  1.38 &      &  1.28 &  1.30 &  1.31 & \multicolumn{2}{c}{1.13} \\
      $\Gamma_{25'v}$ &        $L_{1c}$ &  1.45 &  2.21 &      &  2.12 &  2.12 &  2.13 & 2.04 & 0.06\\
      $\Gamma_{25'v}$ &        $L_{3c}$ &  3.24 &  4.18 &      &  3.91 &  3.93 &  3.94 &  3.9 &  0.1\\
      \hline
      \multicolumn{2}{c}{MSE} & 0.35 &  0.02 &      &  0.01 &  0.00 &  0.03\\
      \multicolumn{2}{c}{MAE} & 0.44 &  0.21 &      &  0.14 &  0.16 &  0.17\\
      \hline
      \hline
   \end{tabular}

   \textsuperscript{\emph{a}} this work;
   \textsuperscript{\emph{b}} Ref.~\citenum{Shishkin2007} for $E_g$ and Ref.~\citenum{Hybertsen1986} for the transitions;
   \textsuperscript{\emph{c}} Ref.~\citenum{Shishkin2007a};
   \textsuperscript{\emph{d}} Ref.~\citenum{DeGennaro2022};
   \textsuperscript{\emph{e}} Ref.~\citenum{Madelung2004}

   \caption{The band gap $E_g$ and energy differences between symmetry points in the band structure of bulk silicon (in eV), calculated with various functional and perturbative approaches. These are compared against experimental values via the mean signed and mean absolute errors (MSE and MAE respectively). The calculated values for the band gap and the valence-to-conduction transitions have been shifted by $-$0.06 eV to account for zero-point renormalization \cite{Miglio2020}.}
   \label{tab:si_band_struct}
\end{table}

\subsection{The band structure of zinc oxide}

\begin{figure}[t]
   \centering
   \includegraphics[width=\columnwidth]{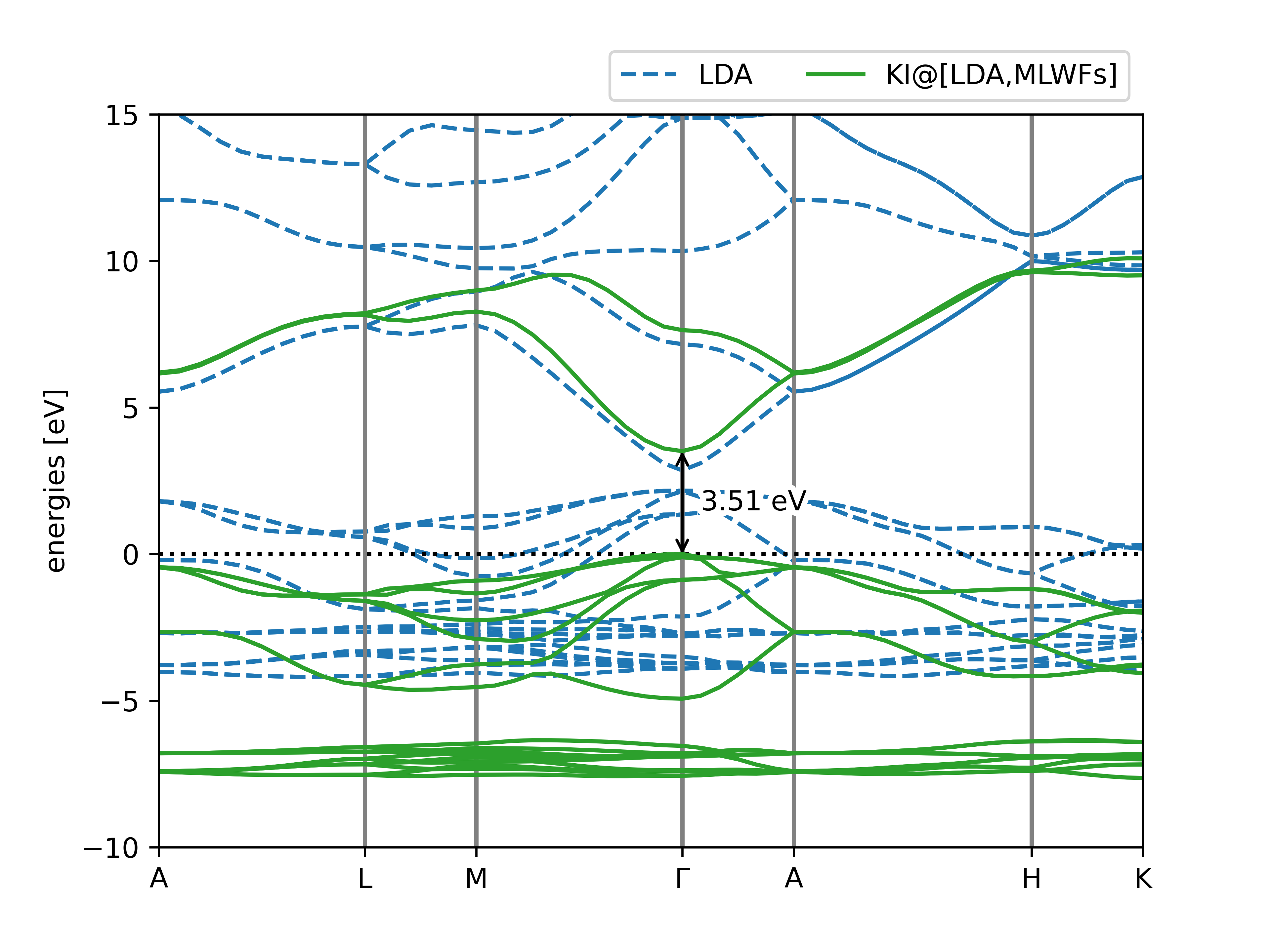}
   \caption{Band structure of zinc oxide, calculated using the KI functional with the LDA base functional, MLWFs as variational orbitals, and screening parameters calculated via DFPT. The LDA band structure is also plotted for comparison.}
   \label{fig:zno_band_struct}
\end{figure}

The previous example used the finite-difference approach for calculating the screening parameters. In this final example, we will instead use the DFPT approach to calculate the band structure of zinc oxide. Refer to Supporting Information~\ref{sec:dfpt_workflow_in_detail} for a step-by-step description of what this entails. The calculated band structure is shown in Figure~\ref{fig:zno_band_struct} and the band gaps are listed in Table~\ref{tab:zno_band_struct}. The corresponding input and output files are provided in Supporting Information~\ref{sec:zno_io_files}. In this instance, the band gap is predicted with better accuracy than state-of-the-art self-consistent GW with vertex corrections. This is also true of the average $d$-band energy, although these bands remain slightly too high in energy relative to experiment. Finally, the bandwidth of the oxygen $2p$ bands (the six highest-energy occupied bands) is much improved going from LDA to KI. Note that this is a major departure from the earlier calculations on silicon, where the KI correction to the occupied bands amounts to a rigid shift and thus such a bandwidth would not change. In this instance, these bands comprise of variational orbitals of multiple different characters, each of which is subject to its own potential shift, and thus the overall band shape can (and does) change.

\begin{table}[h]
   \centering
   %%diff%%\color{blue}
   \begin{tabular}{c *{8}{d{2.2}}}
      \hline
      \hline
                                    & \multicolumn{2}{c}{LDA}
                                    & \multicolumn{1}{c}{\multirow{2}{*}{G\textsubscript{0}W\textsubscript{0}\textsuperscript{\emph{a}}}}
                                    & \multicolumn{1}{c}{\multirow{2}{*}{GW\textsubscript{0}\textsuperscript{\emph{b}}}}
                                    & \multicolumn{1}{c}{\multirow{2}{*}{scG$\tilde{\mathrm{W}}$\textsuperscript{\emph{b}}}}
                                    & \multicolumn{2}{c}{KI@[LDA,MLWFs]}
                                    & \multicolumn{1}{c}{\multirow{2}{*}{exp}}                                                                                                                                                                                                                                                                                       \\
                                    & \multicolumn{1}{c}{this work}                                                                       & \multicolumn{1}{c}{Ref.~\citenum{Colonna2022}}
                                    &                                                                                                     &                                                &      & \multicolumn{1}{c}{this work} & \multicolumn{1}{c}{Ref.~\citenum{Colonna2022}}                                                                                   \\
      \hline
      $E_g$                         & 0.53                                                                                                & 0.63                                           & 1.96 & 2.84                          & 3.04                                           & 3.35  & 3.52  & \multicolumn{1}{c}{3.44\textsuperscript{\emph{c}}}              \\
      $\langle\varepsilon_d\rangle$ & -5.42                                                                                               & -5.14                                          & -6.1 & -6.4                          & -6.7                                           & -7.06 & -6.93 & \multicolumn{1}{c}{$-$7.5 to $-$8.81\textsuperscript{\emph{d}}} \\
      $\Delta$                      & 4.43                                                                                                & 4.15                                           &      &                               &                                                & 4.93  & 4.99  & \multicolumn{1}{c}{5.3\textsuperscript{\emph{e}}}               \\
      \hline
      \hline
   \end{tabular}
   \textsuperscript{\emph{a}} Ref.~\citenum{Shishkin2007}; \textsuperscript{\emph{b}} Ref.~\citenum{Shishkin2007a}; \textsuperscript{\emph{c}} Ref.~\citenum{Kittel2004}; \textsuperscript{\emph{d}} Ref.~\citenum{Ley1974}; \textsuperscript{\emph{e}} Ref.~\citenum{Kobayashi2009}
   \caption{The band gap $E_g$, average $d$-band energy $\langle \varepsilon_d \rangle$, and bandwidth $\Delta$ of bulk zinc oxide (all in eV), as given by various functional and perturbative approaches, and compared to experiment. All of the computational values for the band gap have been shifted by -0.16 eV to account for zero-point renormalization \cite{Manjon2003,Miglio2020}. In contrast to the earlier calculation on bulk silicon, here we used the LDA base functional to align with Ref.~\citenum{Colonna2022}. That work used finer parameters than this work (most notably, a finer $k$-point grid throughout).}
   \label{tab:zno_band_struct}
\end{table}

\section{Conclusions}\label{sec:conclusion}
% Summary
Koopmans functionals are a powerful computational tool for predicting the spectral properties of atoms, molecules, liquids, and crystalline and amorphous solids from first principles with a functional approach. This has already been demonstrated in their ability to calculate the ionization potentials and electron affinities of small molecules,\cite{Nguyen2015,Colonna2019} the photoemission spectra of large molecules\cite{Nguyen2015,Nguyen2016}, the electronic structure of liquid water\cite{deAlmeida2021}, and the band structures and ionization potentials of prototypical semiconductors and insulators\cite{Nguyen2018,DeGennaro2022,Colonna2022}, all at a level of accuracy comparable to state-of-the-art many-body perturbation methods.

The newly released \texttt{koopmans} package now makes it possible, for the first time, for non-experts to use these functionals in their own research. Experts will also benefit from their calculations becoming much more robust and reproducible. For more information, we refer the reader to the website \href{https://koopmans-functionals.org/}{\texttt{koopmans-functionals.org}}.

% Future
The \texttt{koopmans} package will continue to be maintained and developed. In particular, Koopmans calculations on periodic systems require the user to perform a Wannierization of the electronic states, and correctly configuring this calculation can be onerous. In the near future we will add support for automated Wannierization \cite{Qiao2023,Qiao2023a}.

The second focus of ongoing development will be parallelization. Large swathes of the Koopmans workflow (for example, the calculation of screening parameters) are embarrassingly parallel. For example, one could calculate a revised value of the screening parameter for orbital $i$ entirely independently of the calculation of the screening parameter for orbital $j$. (This is true for both the finite difference and DFPT schemes.) However, \texttt{koopmans} performs each calculation in the workflow serially i.e.\ multiple calculations are not run simultaneously. (N.B. We are not saying that individual calculations must be run on a single core; all the codes support MPI parallelization.) Integration of the workflows within a workflow engine such as AiiDA would allow us to massively reduce the workflows' walltimes \cite{Huber2020}. Integration within AiiDA would come with the added benefits of AiiDA's provenance tracking and error detection/recovery. Combined with the automated Wannierization and efficient parallelism, high-throughput studies with Koopmans functionals are just around the corner.

\begin{acknowledgement}

   We gratefully acknowledge financial support from the Swiss National Science Foundation (SNSF -- project numbers 179138 and 213082). This research was also supported by the NCCR MARVEL, a National Centre of Competence in Research, funded by the Swiss National Science Foundation (grant number 205602).  We also acknowledge support from MaX -- MAterials design at the eXascale -- a European Centre of Excellence funded by the European Union's program HORIZON-EUROHPC-JU-2021-COE-01 (grant number 101093374). NLN acknowledges support by the Vietnam National Foundation for Science and Technology Development (NAFOSTED) under grant number 103.02-2021.95. ID acknowledges support from the Division of Materials Research of the National Science Foundation through CAREER Award number DMR-1654625. We thank Susi Lehtola for his helpful comments on the manuscript.

\end{acknowledgement}

\begin{suppinfo}

   The Supporting Information contains a derivation of the functional form of Koopmans functionals, further details on the KIPZ functional, a detailed description of the Koopmans workflows, example input and output \texttt{koopmans} files, and additional details regarding the \texttt{koopmans} package. All of the input and output files related to this paper can also be found on Materials Cloud at \href{https://doi.org/10.24435/materialscloud:9w-sp}{10.24435/materialscloud:9w-sp}.

\end{suppinfo}


\newpage

\input{si_body.tex}

\newpage
\bibliography{references}

\end{document}

%% file: seaborn_colours.tex
\definecolor{seaborn_bg_grey}{HTML}{eaeaf2}
\definecolor{seaborn_bg_grey_dark}{HTML}{d2d2d9}
\definecolor{seaborn_bg_grey_darker}{HTML}{a3a3a9}
\definecolor{seaborn_bg_grey_half}{HTML}{f4f4f8}

\definecolor{seaborn_blue}{HTML}{4c72b0}
\definecolor{seaborn_green}{HTML}{55a868}
\definecolor{seaborn_red}{HTML}{c44e52}
\definecolor{seaborn_magenta}{HTML}{8172b2}
\definecolor{seaborn_yellow}{HTML}{ccb974}
\definecolor{seaborn_cyan}{HTML}{64b5cd}

\definecolor{seaborn_muted_blue}{HTML}{4878cf}
\definecolor{seaborn_muted_green}{HTML}{6acc65}
\definecolor{seaborn_muted_red}{HTML}{d65f5f}
\definecolor{seaborn_muted_magenta}{HTML}{b47cc7}
\definecolor{seaborn_muted_yellow}{HTML}{c4ad66}
\definecolor{seaborn_muted_cyan}{HTML}{77bedb}

\definecolor{seaborn_pastel_blue}{HTML}{92c6ff}
\definecolor{seaborn_pastel_green}{HTML}{97f0aa}
\definecolor{seaborn_pastel_red}{HTML}{ff9f9a}
\definecolor{seaborn_pastel_magenta}{HTML}{d0bbff}
\definecolor{seaborn_pastel_yellow}{HTML}{fffea3}
\definecolor{seaborn_pastel_cyan}{HTML}{b0e0e6}

\definecolor{seaborn_bright_blue}{HTML}{003fff}
\definecolor{seaborn_bright_green}{HTML}{03ed3a}
\definecolor{seaborn_bright_red}{HTML}{e8000b}
\definecolor{seaborn_bright_magenta}{HTML}{8a2be2}
\definecolor{seaborn_bright_yellow}{HTML}{ffc400}
\definecolor{seaborn_bright_cyan}{HTML}{00d7ff}

\definecolor{seaborn_dark_blue}{HTML}{001c7f}
\definecolor{seaborn_dark_green}{HTML}{017517}
\definecolor{seaborn_dark_red}{HTML}{8c0900}
\definecolor{seaborn_dark_magenta}{HTML}{7600a1}
\definecolor{seaborn_dark_yellow}{HTML}{b8860b}
\definecolor{seaborn_dark_cyan}{HTML}{006374}

\definecolor{seaborn_colorblind_blue}{HTML}{0072b2}
\definecolor{seaborn_colorblind_green}{HTML}{009e73}
\definecolor{seaborn_colorblind_red}{HTML}{d55e00}
\definecolor{seaborn_colorblind_magenta}{HTML}{cc79a7}
\definecolor{seaborn_colorblind_yellow}{HTML}{f0e442}
\definecolor{seaborn_colorblind_cyan}{HTML}{56b4e9}

%% file: si_body.tex
\renewcommand{\thetable}{S\arabic{table}}  
\setcounter{table}{0}
\renewcommand{\thefigure}{S\arabic{figure}}
\setcounter{figure}{0}
\renewcommand{\thesection}{S\arabic{section}}
\setcounter{section}{0}
\renewcommand{\theequation}{S\arabic{equation}}
\setcounter{equation}{0}

{\Large \noindent \textbf{Supporting Information}}
\section{Derivation of the functional form of Koopmans functionals}
\label{sec:derivation_of_functional_form}
The brief derivation of the functional form of Koopmans functionals is as follows: let us assume a functional of the form
\begin{equation}
   E^\text{Koopmans} = E^\mathrm{DFT} + \sum_i \Pi_i
\end{equation}
If we take the derivative with respect to the occupancy of the $j^{\rm th}$ variational orbital then we have
\begin{equation}
   \eta_j = \left.\frac{dE^{DFT}}{df_j}\right\rvert_{f_j=f} + \left.\frac{d\Pi_j}{df_j}\right\rvert_{f_j=f} = \langle \varphi_j\rvert \hat h^\text{DFT}(f)\lvert\varphi_j\rangle
   + \left.\frac{d\Pi_j}{df_j}\right\rvert_{f_j=f}
\end{equation}
where we assumed that the cross-term derivatives $d\Pi_i/df_j$ vanish, and because $E^\text{Koopmans}$ ought to be linear in $f_j$, we replaced its derivative with some yet-to-be deterimined constant $\eta_j$. For the second equality we invoked Janak's theorem, and $f$ is some number between 0 and 1.

Assuming that the energy correction $\Pi_j$ is zero at integer occupancies, is independent of $f_i$ for $i \neq j$, and neglecting for the moment any orbital relaxation as the orbital occupancies change, it follows that
\begin{equation}
   \Pi^u_j =  - \int_0^{f_j} \langle \varphi_j\rvert \hat h^\text{DFT}(f)\lvert\varphi_j\rangle df + f_j \eta_j
   = - \left(
   E^\mathrm{DFT}[\rho] - E^\mathrm{DFT}[\rho - \rho_i]
   \right)
   + f_j \eta_j
\end{equation}
where the $u$ superscript denotes the fact that we neglected orbital relaxation, and thus this term is ``unscreened''. To account for this screening we must introduce some screening parameters $\{\alpha_i\}$ such that $\Pi_j = \alpha_j \Pi^u_j$. Having done this, we arrive at eq.~\ref{eqn:Koopmans_functional_form}, the final result.

\section{KIPZ details}
\label{sec:kipz_details}

In previous works, KIPZ has been presented in slightly different ways. In eq.~27 of Ref.~\citenum{Borghi2014}, KIPZ was introduced as
\begin{align}
\Pi_i^{\mathrm{KIPZ}}= & -\int_0^{f_i}\langle\varphi_i|\hat{H}^{\mathrm{PZ}}_i(s)| \varphi_i\rangle d s +f_i \int_0^1\langle\varphi_i|\hat{H}^{\mathrm{PZ}}_i(s)| \varphi_i\rangle d s-E_{\mathrm{Hxc}}[\rho_i]\label{eqn:borghi27}
\end{align}
where  $\hat{H}_i^{\mathrm{PZ}}(s)=\hat{H}^{\mathrm{DFT}}(s)-\hat{v}_{\mathrm{Hxc}}^{\mathrm{DFT}}[s|\varphi_i(\mathbf{r})|^2]$. (Ref.~\citenum{Borghi2014} included an erroneous sum over $i$ in the definition of this Hamiltonian.)
Meanwhile, eq.~6 of Ref~\citenum{Nguyen2018} defined KIPZ as
\begin{align}
\Pi_i^{\mathrm{KIPZ}}= & -\int_0^{f_i}\langle\varphi_i|\hat{H}^{\mathrm{DFT}}(s)| \varphi_i\rangle d s 
+f_i \int_0^1\langle\varphi_i|\hat{H}_i^{\mathrm{PZ}}(s)| \varphi_i\rangle d s\label{eqn:nguyen6}
\end{align}
and in that same paper it was also stated that
\begin{align} \Pi_i^\mathrm{KIPZ}=& \Pi_i^\mathrm{KI} - f_i E_\mathrm{Hxc}[n_i]\label{eqn:nguyen8}
\end{align}
One can prove that these three definitions are equivalent via the identity
\begin{align}
& \int_0^f\langle\varphi_i| v_{\mathrm{Hxc}}[s n_i]| \varphi_i\rangle d s = E_\mathrm{Hxc}[f n_i],
\end{align}
from which it follows that 
\begin{align}
&  \int_0^{f_i}\langle\varphi_i|\hat{H}^{\mathrm{PZ}}_i(s)| \varphi_i\rangle d s \nonumber \\
= & \int_0^{f_i}\langle\varphi_i|\hat{H}^{\mathrm{DFT}}(s)| \varphi_i\rangle d s
-  \int_0^{f_i}\langle\varphi_i|v_{\mathrm{Hxc}}[s n_i]| \varphi_i\rangle d s
\nonumber \\
= & \int_0^{f_i}\langle\varphi_i|\hat{H}^{\mathrm{DFT}}(s)| \varphi_i\rangle d s
- E_\mathrm{Hxc}[f_i n_i]
\end{align}
which proves that eqs.~\ref{eqn:borghi27} and \ref{eqn:nguyen6} are equivalent. Furthermore,
\begin{align}
& -\int_0^{f_i}\langle\varphi_i|\hat{H}^{\mathrm{DFT}}(s)| \varphi_i\rangle d s 
+f_i \int_0^1\langle\varphi_i|\hat{H}_i^{\mathrm{PZ}}(s)| \varphi_i\rangle d s \nonumber \\
= & -\int_0^{f_i}\langle\varphi_i|\hat{H}^{\mathrm{DFT}}(s)| \varphi_i\rangle d s 
+f_i \int_0^1\langle\varphi_i|\hat{H}_i^{\mathrm{DFT}}(s)| \varphi_i\rangle d s
- f_i \int_0^1\langle\varphi_i| v_{\mathrm{Hxc}}[s n_i]| \varphi_i\rangle d s \nonumber \\
= & \Pi_i^\mathrm{KI}
- f_i E_\mathrm{Hxc}[n_i]
\end{align}
and thus eqs.~\ref{eqn:nguyen6} and \ref{eqn:nguyen8} are equivalent.

   In the unscreened case, the KIPZ functional as defined above is equivalent to the KI correction applied to an unscreened PZ base functional (i.e.\ KI@PZ). However, in the general case when screening is accounted for, KIPZ and KI@PZ are not equivalent. Instead, the PZ corrections incorporated within the KIPZ functional each inherit their own screening coefficient from the generalized PWL condition. This is desirable, because scaling down the PZ correction has been shown to improve energetics and thermochemistry \cite{Vydrov2006,Bylaska2006,Valdes2011}, but it would be interesting to explore alternative prescriptions for the scaling of the PZ correction that are decoupled from the generalized PWL condition.

\begin{landscape}
\section{Details of the Koopmans workflows}
\label{sec:workflows_in_detail}

This appendix contains a detailed breakdown of the two key Koopmans workflows: one for calculations where the screening parameters are calculated via finite differences (Figure~\ref{fig:dscf_workflow}), and the other via DFPT (Figure~\ref{fig:dfpt_workflow})

\subsection{The finite-difference workflow}
\label{sec:dscf_workflow_in_detail}
In this workflow, we calculate screening parameters via the method described in Section~\ref{sec:calculating_alphas} of the main text.

\begin{figure}[h!]
   \centering
   \adjustbox{width=\columnwidth}{\input{supercell_workflow.tex}}
   \caption{The finite-difference workflow. Individual nodes represent a calculation performed with \texttt{\textsc{Quantum ESPRESSO}}, except for the red nodes, which represent key quantities of interest that we extract from a preceding calculation. Individual nodes are explained more fully in the text.}
   \label{fig:dscf_workflow}
\end{figure}
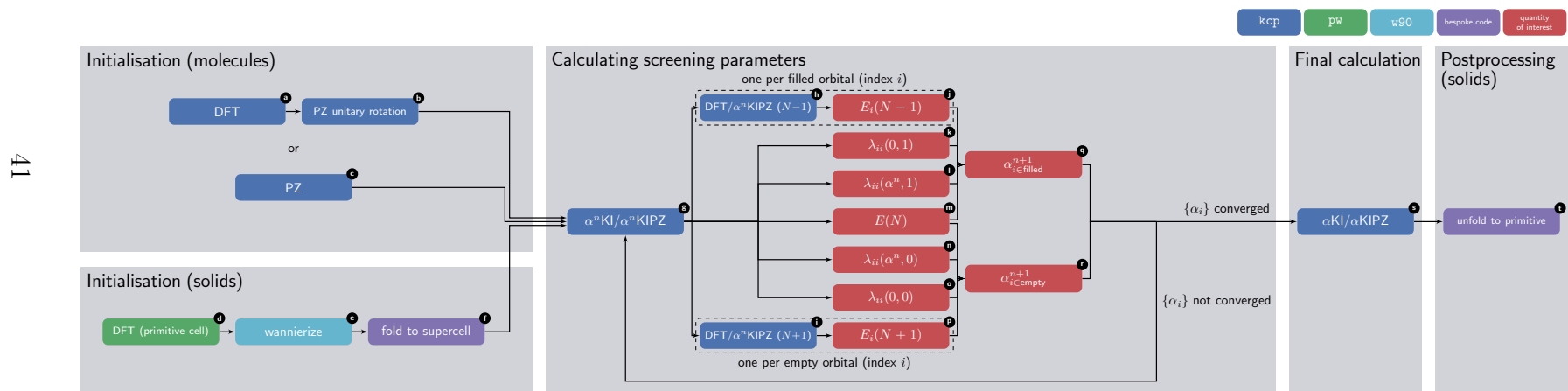

\end{landscape}

\subsubsection{Initialization}
\label{sec:workflows_in_detail_init}
The first step in this workflow is the initialization of the density and the variational orbitals. Depending on the system and functional in question, this can look quite different.  For molecules, one can start with a DFT calculation to obtain the ground-state Kohn-Sham eigenvalues (node a in Figure~\ref{fig:dscf_workflow}). This is typical of KI calculations, which share the same ground-state density as the base DFT functional, in which case the final density has already been determined by this very first calculation. Note, however, that a unitary rotation of the occupied Kohn-Sham orbital densities leaves the total density (and therefore the total energy) unchanged, which means that the variational orbitals from a DFT calculation are not uniquely defined. In order to resolve this issue, one then performs a unitary rotation of the occupied Kohn-Sham variational orbitals to minimize the PZ energy (node b). This gives us a unique set of variational orbitals, while leaving the total density unchanged, and fulfils the definition of the KI functional as the $\gamma \rightarrow 0$ limit of the ``KI$\gamma$PZ" functional (as introduced in Section~\ref{sec:pkipz}) in the main text.  If one is using the KIPZ functional, it is better to initialize the density and the variational orbitals by performing a full PZ calculation (node c). In contrast to the previous KI approach where the DFT and KI ground-state densities match, the KIPZ and PZ ground-state densities and variational orbitals are similar but not identical, so the PZ solution serves as a suitable initial guess for a KIPZ calculation.  Note that all of the above calculations are performed with the $\Gamma$-point-only \texttt{kcp.x} code.

For solids, the approach for initializing the density and variational orbitals is very different. Here, we take advantage of the periodicity of the lattice by initializing the variational orbitals using maximally localized Wannier functions (MLWFs). This approach is justified by the Wannier-like character of the true minimizing orbitals \cite{DeGennaro2022}. Practically, the Wannierization procedure involves a DFT calculation with the \texttt{pw.x} code (node d), followed by a Wannierization procedure using the \texttt{Wannier90} and \texttt{pw2wannier90.x} codes (node e).

There are two important points when it comes to the Wannierization. The first is that the occupied and the empty manifolds must be Wannierized separately. This guarantees that the occupancy matrix is diagonal in the basis of variational orbitals, as required by Koopmans functionals (see Section~\ref{sec:other_considerations} of the main text). The second important point is that mixing bands that are far apart in energy-space is generally detrimental to the Koopmans results. To avoid this, each block of bands that are well-separated in energy-space are Wannierized separately, preventing inter-block mixing during the Wannierization procedure. This is a similar but cruder approach to the so-called dually-localized Wannier functions, where the Wannier functions minimize a localization criteria that is a mix of spatial and energy localization \cite{Mahler2022}.

The one final task is to map the Wannier functions in the $k$-sampled primitive cell to the equivalent $\Gamma$-point-only supercell in a format readable by the \texttt{kcp.x} code that will handle the subsequent calculation of the screening parameters (node f). In this procedure the supercell dimensions match those of the $k$-grid used during the initialization.

\subsubsection{Calculating the screening parameters}

Having initialized the density and the variational orbitals, the next task to perform is the calculation of the screening parameters. To this end, let us restate equations~\ref{eqn:alpha_via_dscf_filled} and \ref{eqn:alpha_via_dscf_empty} from the main text, with which we calculate these parameters:
\begin{equation}
   \alpha^{n+1}_i =
   \alpha^n_i \frac{\Delta E^\text{Koopmans}_i - \lambda_{ii}(0, 1)}{\lambda_{ii}(\alpha^n_i, 1) - \lambda_{ii}(0, 1)}; \qquad  \Delta E^\text{Koopmans}_i = E^\text{Koopmans}(N) - E^\text{Koopmans}_i(N - 1)
\end{equation}
for occupied orbitals and
\begin{equation}
   \alpha^{n+1}_i =
   \alpha^n_i \frac{\Delta E^\text{Koopmans}_i - \lambda_{ii}(0,0)}{\lambda_{ii}(\alpha^n_i,0) - \lambda_{ii}(0,0)}; \qquad \Delta E^\text{Koopmans}_i = E^\text{Koopmans}_i(N+1) - E^\text{Koopmans}(N)
\end{equation}
for empty orbitals. In order to calculate these screening parameters, we therefore require three calculations. The first calculation is a KI or KIPZ calculation with using a trial screening parameter $\alpha^0$ (node g). This gives us access to the energy of the $N$-electron system $E(N)$ (node m) as well as all of the requisite expectation values Koopmans Hamiltonian on the variational orbitals $\lambda_{ii}(\alpha, f)$ (nodes k, l, n, and o). The second and third calculations (nodes h and i) are calculations on the $N\pm1$-electron systems, where orbital $i$ is frozen and its occupancy is fixed to 0 (in the case of occupied orbitals) or 1 (empty orbitals). These calculations yield the total energies $E_i(N\pm 1)$ (nodes j and p). For these calculations in particular, ensuring that there is no spurious interactions between images is crucial (because now we have a charged defect in our system). This requires the use of both a sufficiently large supercell and a correction scheme such as Gygi-Baldereschi \cite{Gygi1986}. Note that since KI yields the same total energies as the base functional, these two calculations can be performed at the DFT level when performing the KI workflow.

Having performed these three calculations (nodes g-i) and extracted all of the requisite information (nodes j-p), we can then calculate the screening parameters (nodes q and r) according to the above equations. If the screening parameters are converged, we can then proceed to the final calculation; if not, the process is repeated.

We note that this iterative procedure almost universally converges very quickly. Indeed, for the KI functional and with occupied orbitals, it is guaranteed to converge instantly. This is because, as mentioned earlier, $\Delta E_i$ is independent of $\alpha$, as are occupied variational orbitals, and consequently $\lambda^\mathrm{KI}_{ii}(\alpha, 1)$ is linear in $\alpha$. This is not the case for empty orbitals for the KI functional (for which $\lambda^\mathrm{KI}_{ii}(\alpha, 0)$ is not strictly linear) or for the KIPZ functional (where additionally $\Delta E_i$ is dependent on $\alpha$). Even for these functionals, the screening parameters tend to converge in a few iterations.

\subsubsection{The final calculation and postprocessing for solids}
We now perform a KI or KIPZ calculation with the finalized set of screening parameters (node s). For a molecular system we are now done: the KI/KIPZ calculation yields a Hamiltonian in the basis of variational orbitals which we diagonalize to extract the quasiparticle energies.

However, for a calculation on a periodic system one final step is required. This is because all of the preceding \texttt{kcp.x} calculations have been performed in a $\Gamma$-point-only supercell. In order to extract the band structure, we must now unfold the band structure by taking advantage of the MLWF basis, as described in Ref.~\citenum{DeGennaro2022}. This step is performed within python by the \texttt{koopmans} workflow manager itself. One trick that we can perform at this stage is ``smooth interpolation''. In the supercell, our Hamiltonian in the basis of Wannier functions is given by
\begin{equation}
   h^{\mathrm{DFT}}_{mn}(\mathbf{R}) + v^{\mathrm{Koopmans}}_{mn}(\mathbf{R})
\end{equation}
The Koopmans potential is very smooth and slowly-varying in $\mathbf{k}$-space, applying an almost-constant shift to the Kohn-Sham DFT bands. Consequently, the dominant contribution to the dispersion of the bands comes from the DFT Hamiltonian, and it makes sense to construct the $\mathbf{k}$-indexed Hamitonian as
\begin{equation}
   h_{mn}(\mathbf{k}) = \sum_{\mathbf{R}'} e^{i\mathbf{k}\cdot \mathbf{R}'} h_{mn}^\mathrm{DFT}(\mathbf{R'})
   +
   \sum_{\mathbf{R}} e^{i\mathbf{k}\cdot \mathbf{R}} v_{mn}^\mathrm{Koopmans}(\mathbf{R})
\end{equation}
where now $\{\mathbf{R}'\}$ corresponds to a much larger supercell or, equivalently, a much denser $\mathbf{k}$-point grid \cite{}. The advantage of this strategy is that it improves the interpolation of the band structure at very little computational cost. Suppose we perform a smooth interpolation with a grid twice as fine as the default grid. The only additional computational cost in this instance is having to generate the DFT Hamiltonian in the Wannier basis for this finer grid (i.e. we repeat nodes d and e). This only represents a small fraction of the total workflow, and includes only DFT and not ODDFT calculations, so it only fractionally increases the total computational cost. Contrast this with the alternative, where one could perform the entire calculation with a grid twice as fine. This would require us to perform (among other things) calculations on a supercell containing eight times as many atoms, drastically increasing the computational cost of the workflow as a whole. For more details on the smooth interpolation procedure, refer to Ref.~\citenum{DeGennaro2022}.

\subsection{The DFPT workflow}
\label{sec:dfpt_workflow_in_detail}

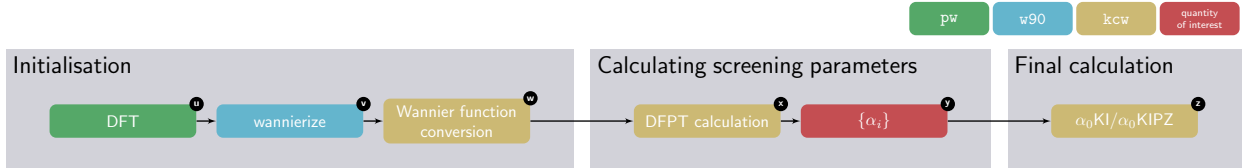
\begin{figure}[t]
   \adjustbox{width=\textwidth}{\input{primitive_workflow.tex}}
   \caption{The DFPT workflow. Individual nodes are explained in the text.}
   \label{fig:dfpt_workflow}
\end{figure}

The DFPT workflow is depicted in Figure~\ref{fig:dfpt_workflow}. It is simpler than the finite-difference procedure, because orbital relaxation is not implemented, and instead Wannier functions are used as approximations to the true variational orbitals. This means that only the KI and pKIPZ functionals can be used in this scheme, means that the screening parameters do not need to be calculated self-consistently, and makes the Koopmans functional effectively a post-processing step on top of a DFT calculation.

\subsubsection{Initialization}
The initialization procedure for the DFPT workflow is very similar to that for solids in the finite-difference workflow. A primitive cell calculation \texttt{pw.x} calculation (node u) is followed by a Wannierization procedure in order to define the density and the variational orbitals (node v). The one difference is that now, instead of using \texttt{kcw.x} to map these Wannier functions to a supercell that is readable by \texttt{kcp.x}, we use \texttt{kcw.x} to convert the Wannier functions to more convenient format for subsequent calculations (node w). Note that \texttt{kcw.x} does not map to a supercell because this workflow operates entirely within the primitive cell with $k$-point sampling.

\subsubsection{Calculating the screening parameters}

DFPT calculations evaluating equation~\ref{eqn:alpha_via_dfpt} (from the main text) are then performed by a single \texttt{kcw.x} run (node x). These calculations yield the screening parameters (node y).

\subsubsection{The final calculation and postprocessing for solids}
Having calculated the screening parameters, the Koopmans Hamiltonian is constructed in the basis of Wannier functions and then diagonalized, resulting in the full band structure of the system at hand (node z). Like the previous calculations, this is performed using \texttt{kcw.x}.

\section{Example input and output files}
\label{sec:io_files}
Below are example \texttt{koopmans} input and output files for ozone, silicon, and zinc oxide. All of these files can also be found on Materials Cloud at \href{https://doi.org/10.24435/materialscloud:9w-sp}{10.24435/materialscloud:9w-sp}.

\subsection{Ozone}
\label{sec:ozone_io_files}
An example minimal input file for ozone is as follows

\begin{minted}{json}
{
  "workflow": {
    "functional": "ki",
    "method": "dscf",
    "init_orbitals": "kohn-sham",
    "n_max_sc_steps": 5,
    "keep_tmpdirs": false,
    "pseudo_library": "sg15"
  },
  "atoms": {
    "cell_parameters": {
      "periodic": false,
      "vectors": [[14.1738, 0.0, 0.0],
                  [0.0, 12.0, 0.0],
                  [0.0, 0.0, 12.66]],
      "units": "angstrom"
    },
    "atomic_positions": {
      "positions": [["O", 7.0869, 6.0, 5.89],
                    ["O", 8.1738, 6.0, 6.55],
                    ["O", 6.0, 6.0, 6.55]],
      "units": "angstrom"
    }
  },
  "calculator_parameters": {
    "ecutwfc": 65.0,
    "nbnd": 10
  }
}
\end{minted}
This input file contains several blocks. The \texttt{workflow} block allows the user to specify the details of the workflow. Here we can see we are performing a KI calculation (line 3) calculating the screening parameters via the finite-difference procedure (line 4), and using the Kohn-Sham orbitals to initialize our variational orbitals (line 5; this is common practice for molecules). The \texttt{atoms} block (lines 10-23) contains standard keywords specifying the system configuration, such as the \texttt{cell\_parameters} and \texttt{atomic\_positions}. These mirror the equivalent blocks in \texttt{Quantum\ ESPRESSO} input files (albeit in JSON format). Finally, the \texttt{calculator\_parameters} block allows the user to specify settings specific to a particular code (e.g. a \texttt{w90} subblock for specifying \texttt{Wannier90} settings). In this instance we are providing a particular energy cutoff (line 26) and specifying the total number of orbitals to compute (line 27).

The output of \texttt{koopmans ozone.json}, which prompts a sequence of \texttt{Quantum ESPRESSO} calculations necessary to initialize the density and variational orbitals (lines 16-22), calculate the screening parameters (lines 24-68), and run a final KI calculation (lines 85-87).

\setminted{fontsize=\footnotesize}
\begin{minted}{text}
  _
 | | _____   ___  _ __  _ __ ___   __ _ _ __  ___
 | |/ / _ \ / _ \| '_ \| '_ ` _ \ / _` | '_ \/ __|
 |   < (_) | (_) | |_) | | | | | | (_| | | | \__ \ 
 |_|\_\___/ \___/| .__/|_| |_| |_|\__,_|_| |_|___/
                 |_|

 Koopmans spectral functional calculations with Quantum ESPRESSO
\end{minted}
\texttt{...}
\begin{minted}{text}
  Initialization of density and variational orbitals
  ==================================================
   Running init/dft_init_nspin1... done
   Running init/dft_init_nspin2_dummy... done
   Running init/dft_init_nspin2... done
   Overwriting the variational orbitals with Kohn-Sham orbitals
   Copying the spin-up variational orbitals over to the spin-down channel

  Calculating screening parameters
  ================================

  SC iteration 1
  --------------
   Running calc_alpha/iteration_1/ki... done

  Orbital 1
  ---------
   Running calc_alpha/iteration_1/orbital_1/dft_n-1... done
\end{minted}
\texttt{...}
\begin{minted}{text}
  Orbital 10
  ----------
   Running calc_alpha/iteration_1/orbital_10/pz_print... done
   Running calc_alpha/iteration_1/orbital_10/dft_n+1_dummy... done
   Running calc_alpha/iteration_1/orbital_10/dft_n+1... done
   
\end{minted}
\texttt{...}
\begin{minted}{text}
  SC iteration 2
  --------------
   Running calc_alpha/iteration_2/ki_nspin1_dummy... done
   Running calc_alpha/iteration_2/ki_nspin1... done
   Running calc_alpha/iteration_2/ki_nspin2_dummy... done
   Running calc_alpha/iteration_2/ki_nspin2... done

  Orbital 10
  ----------
   Running calc_alpha/iteration_2/orbital_10/pz_print... done
   Running calc_alpha/iteration_2/orbital_10/dft_n+1_dummy... done
   Running calc_alpha/iteration_2/orbital_10/dft_n+1... done
   
   alpha
            1         2         3        4   ...        7       8         9         10
   0  0.600000  0.600000  0.600000  0.60000  ...  0.600000  0.6000  0.600000  0.600000
   1  0.655689  0.727571  0.783859  0.66386  ...  0.729888  0.7419  0.779264  0.717389
   2  0.655689  0.727571  0.783859  0.66386  ...  0.729888  0.7419  0.779264  0.717389
   
\end{minted}
\texttt{...}
\begin{minted}{text}
   Screening parameters have been converged

  Final KI calculation
  ====================
   Running final/ki_final... done

 Workflow complete
\end{minted}

\subsection{Silicon}
\label{sec:silicon_io_files}

A typical input file for silicon is similar to that of ozone, but with some tweaked workflow settings, a \texttt{cell\_parameters} block that denotes that this system is periodic, and some additional Wannierization settings:

\begin{minted}{json}
{
  "workflow": {
    "functional": "ki",
    "base_functional": "pbe",
    "method": "dscf",
    "init_orbitals": "mlwfs",
    "alpha_guess": 0.1,
    "eps_inf": 13.02,
    "orbital_groups_self_hartree_tol": 0.1,
    "pseudo_library": "sg15",
    "mp_correction": false
  },
  "atoms": {
    "atomic_positions": {
      "units": "crystal",
      "positions": [["Si", 0.00, 0.00, 0.00],
              ["Si", 0.25, 0.25, 0.25]]
    },
    "cell_parameters": {
      "periodic": true,
      "ibrav": 2,
      "celldms": {"1": 10.2622}
    }
  },
  "kpoints": {
    "grid": [4, 4, 4]
  },
  "calculator_parameters": {
    "ecutwfc": 60.0,
    "pw": {
      "nbnd": 20
    },
    "w90": {
      "projections": [
        [{"fsite": [0.25, 0.25, 0.25], "ang_mtm": "sp3"}],
        [{"fsite": [0.25, 0.25, 0.25], "ang_mtm": "sp3"}]],
      "dis_froz_max": 11.0,
      "dis_win_max": 16.5
    },
    "ui": {
      "smooth_int_factor": 2
    }
  }
}
\end{minted}

For a full explanation of the meaning of the \texttt{Wannier90} keywords we refer the reader to the \texttt{Wannier90} documentation.

Running \texttt{koopmans si.json} generates the following output. This command prompts a sequence of \texttt{Quantum ESPRESSO} calculations necessary to initialize the density and variational orbitals using Wannier functions (lines 20-29), fold these $k$-resolved functions to the equivalent $\Gamma$-only supercell (lines 31-36) calculate the screening parameters (lines 38-73), run a final KI calculation (lines 75-77), and finally perform a second DFT Wannierization on a finer $k$-grid to produce a smoothly interpolated band structure (lines 80-97).

\begin{minted}{text}
  _
 | | _____   ___  _ __  _ __ ___   __ _ _ __  ___
 | |/ / _ \ / _ \| '_ \| '_ ` _ \ / _` | '_ \/ __|
 |   < (_) | (_) | |_) | | | | | | (_| | | | \__ \ 
 |_|\_\___/ \___/| .__/|_| |_| |_|\__,_|_| |_|___/
                 |_|

 Koopmans spectral functional calculations with Quantum ESPRESSO
\end{minted}
\texttt{...}
\begin{minted}{text}
  Initialization of density and variational orbitals
  ==================================================

   Wannierization
   ==============
    Running wannier/scf... done
    Running wannier/nscf... done
    Running wannier/block_1/wann_preproc... done
    Running wannier/block_1/pw2wan... done
    Running wannier/block_1/wann... done
    Running wannier/block_2/wann_preproc... done
    Running wannier/block_2/pw2wan... done
    Running wannier/block_2/wann... done

   Folding to supercell
   --------------------
    Running block_1/w2kcp... done
    Running block_2/w2kcp... done
   Running init/dft_dummy... done
   Running init/dft_init... done

  Calculating screening parameters
  ================================
   Running calc_alpha/ki... done

  Orbitals 1-255
  --------------
   Skipping; will use the screening parameter of an equivalent orbital

  Orbital 256
  -----------
   Running calc_alpha/orbital_256/dft_n-1... done

  Orbital 257
  -----------
   Running calc_alpha/orbital_257/pz_print... done
   Running calc_alpha/orbital_257/dft_n+1_dummy... done
   Running calc_alpha/orbital_257/dft_n+1... done

  Orbitals 258-512
  ----------------
   Skipping; will use the screening parameter of an equivalent orbital
   
   alpha
           1         2         3    ...       510       511       512
   0  0.100000  0.100000  0.100000  ...  0.100000  0.100000  0.100000
   1  0.104648  0.104648  0.104648  ...  0.047709  0.047709  0.047709
   
   [2 rows x 512 columns]
   
   Delta E_i - epsilon_i (eV)
           1         2         3    ...       510       511       512
   0 -0.023309 -0.023309 -0.023309  ... -0.139624 -0.139624 -0.139624
   
   [1 rows x 512 columns]

   Screening parameters have been determined but are not necessarily converged

  Final KI calculation
  ====================
   Running final/ki_final... done

  Postprocessing
  ===============

    Wannierization
    ==============
     Running wannier/scf... done
     Running wannier/nscf... done
     Running wannier/block_1/wann_preproc... done
     Running wannier/block_1/pw2wan... done
     Running wannier/block_1/wann... done
     Running wannier/block_2/wann_preproc... done
     Running wannier/block_2/pw2wan... done
     Running wannier/block_2/wann... done
     Running wannier/bands... done
UserWarning: Some of the pseudopotentials do not have PP_PSWFC blocks, which means a projected DOS calculation is not possible. Skipping...
    Running occ/ki... done
    Running emp/ki... done
UserWarning: The DOS will not be plotted, because the Brillouin zone is too poorly sampled for the specified value of smearing. In order to generate a DOS, increase the k-point density ("kpath_density" in the "setup" "k_points" subblock) and/or the smearing ("degauss" in the "plot" block)

 Workflow complete
\end{minted}
\newpage
\subsection{Zinc oxide}
\label{sec:zno_io_files}

The input file is as follows:

\begin{minted}{json}
{
    "workflow": {
        "task": "singlepoint",
        "functional": "ki",
        "base_functional": "lda",
        "method": "dfpt",
        "init_orbitals": "mlwfs",
        "calculate_alpha" : true,
        "pseudo_library": "pseudo_dojo_standard",
        "gb_correction" : true,
        "eps_inf": 5.3,
        "from_scratch": true,
        "npool": 4,
        "dfpt_coarse_grid": [2, 2, 2],
        "orbital_groups_spread_tol": 0.0005
    },
    "atoms": {
       "cell_parameters": {
         "periodic": true,
         "ibrav": 4,
         "celldms": {"1": 6.14057, "3": 1.60204}
        },
        "atomic_positions": {
            "units": "crystal",
            "positions": [
                ["Zn", 0.33330, 0.66670, 0.50000 ],
                ["Zn", 0.66670, 0.33330, 0.00000 ],
                ["O" , 0.33330, 0.66670, 0.11725 ],
                ["O" , 0.66670, 0.33330, 0.61725 ]
            ]
        }
    },
    "kpoints": {
        "grid": [
            4,
            4,
            4
        ],
        "path": "ALMGAHK"
    },
    "calculator_parameters": {
        "ecutwfc": 50.0,
        "pw": {
            "system": {
                "nbnd": 52
            }
        },
        "w90": {
            "conv_tol": 1e-8,
            "projections": [
                [{"site": "Zn", "ang_mtm": "l=0"}],
                [{"site": "Zn", "ang_mtm": "l=1"}],
                [{"site": "O", "ang_mtm": "l=0"}],
                [{"site": "Zn", "ang_mtm": "l=2"},
                 {"site": "O", "ang_mtm": "l=1"}],
                [{"site": "Zn", "ang_mtm": "l=0"}]
            ],
            "dis_froz_max": 14.5,
            "dis_win_max": 17.0
        }
    },
    "plotting": {
        "Emin": -10,
        "Emax": 10
    }
}
\end{minted}

Here we can see the selection of DFPT for calculating screening parameters (line 6), the choice of MLWFs as the variational orbitals (line 7), and a criterion for grouping variational orbitals together based on their spreads (line 15). We also have specified a coarse $2\times 2 \times 2$ $k$-point grid on which to calculate the screening parameters (line 14) relative to the $4\times 4 \times 4$ grid upon which we construct the Hamiltonian (specified elsewhere in the output file). Again, it is worth stressing that these calculations are not fully converged.

The output of \texttt{koopmans zno.json} is below. This command prompts a sequence of \texttt{Quantum ESPRESSO} calculations which, using a coarse grid, calculates the density and variational orbitals using Wannier functions (lines 20-40) and then calculates the screening parameters using DFPT (lines 46-57). Then, on a regular grid, it repeats a Wannierization (62-82) and then constructs and diagonalizes the Koopmans Hamiltonian (lines 88-90), using the screening parameters calculated on the coarse grid.

\begin{minted}{text}
  _
 | | _____   ___  _ __  _ __ ___   __ _ _ __  ___
 | |/ / _ \ / _ \| '_ \| '_ ` _ \ / _` | '_ \/ __|
 |   < (_) | (_) | |_) | | | | | | (_| | | | \__ \ 
 |_|\_\___/ \___/| .__/|_| |_| |_|\__,_|_| |_|___/
                 |_|

 Koopmans spectral functional calculations with Quantum ESPRESSO
\end{minted}
\texttt{...}
\begin{minted}{text}
  Coarse grid calculations
  ========================

    Wannierization
    ==============
     Running wannier/scf... done
     Running wannier/nscf... done
     Running wannier/block_1/wann_preproc... done
     Running wannier/block_1/pw2wan... done
     Running wannier/block_1/wann... done
     Running wannier/block_2/wann_preproc... done
     Running wannier/block_2/pw2wan... done
     Running wannier/block_2/wann... done
     Running wannier/block_3/wann_preproc... done
     Running wannier/block_3/pw2wan... done
     Running wannier/block_3/wann... done
     Running wannier/block_4/wann_preproc... done
     Running wannier/block_4/pw2wan... done
     Running wannier/block_4/wann... done
     Running wannier/block_5/wann_preproc... done
     Running wannier/block_5/pw2wan... done
     Running wannier/block_5/wann... done
     Running wannier/bands... done
     Running pdos/projwfc... done

   Conversion to Koopmans format
   -----------------------------
    Running wannier/kc... done

   Calculation of screening parameters
   ===================================
    Running screening/band_2/kc... done
    Running screening/band_6/kc... done
    Running screening/band_8/kc... done
    Running screening/band_10/kc... done
    Running screening/band_16/kc... done
    Running screening/band_18/kc... done
    Running screening/band_20/kc... done
    Running screening/band_24/kc... done
    Running screening/band_26/kc... done
    Running screening/band_27/kc... done

  Regular grid calculations
  =========================

   Wannierization
   ==============
    Running wannier/scf... done
    Running wannier/nscf... done
    Running wannier/block_1/wann_preproc... done
    Running wannier/block_1/pw2wan... done
    Running wannier/block_1/wann... done
    Running wannier/block_2/wann_preproc... done
    Running wannier/block_2/pw2wan... done
    Running wannier/block_2/wann... done
    Running wannier/block_3/wann_preproc... done
    Running wannier/block_3/pw2wan... done
    Running wannier/block_3/wann... done
    Running wannier/block_4/wann_preproc... done
    Running wannier/block_4/pw2wan... done
    Running wannier/block_4/wann... done
    Running wannier/block_5/wann_preproc... done
    Running wannier/block_5/pw2wan... done
    Running wannier/block_5/wann... done
    Running wannier/bands... done
    Running pdos/projwfc... done

  Conversion to Koopmans format
  -----------------------------
   Running wannier/kc... done

  Construction of the Hamiltonian
  ===============================
   Running hamiltonian/kc... done

 Workflow complete
\end{minted}

\section{Details of the \texttt{koopmans} package}
\label{sec:koopmans_code_details}
\subsection{Code structure}
\texttt{koopmans} is built on top of the \texttt{ASE} python package (the Atomic Simulation Environment) \cite{Larsen2017}. Under the hood, it defines various \texttt{Workflow} classes, which look like

\begin{minted}[baselinestretch=1]{python}
class Workflow:
   parameters: Dict[str, Any]
   calculations: List[Calculator]
   ...
\end{minted}
where the \texttt{parameters} attribute is a dictionary that stores the workflow parameters as specified in the input file, and \texttt{calculations} is a list of the calculations in the workflow. The individual entries in the \texttt{calculations} list correspond to \texttt{Calculator} objects:

\begin{minted}[baselinestretch=1]{python}
class Calculator(ASE_Calculator):
   atoms: Atoms
   parameters: Dict[str, Any]
   results: Dict[str, Any]
   ...
\end{minted}
which are subclasses of corresponding classes defined by \texttt{ASE}. \texttt{ASE} provides the calculator with the functionality to read and write input and output files (among many other things).

A \texttt{Calculator} object has --- among others --- an \texttt{atoms} attribute that stores the details of the atoms and the simulation cell. The \texttt{atoms} attribute is itself an instance of the \texttt{Atoms} class from \texttt{ASE}. We note that this hierarchy (namely, that the \texttt{atoms} object is an attribute of a \texttt{Calculator}, and not the other way around) is the reverse of the philosophy of \texttt{ASE}, where \texttt{Atoms} objects are the principal object, and they may or may not have an associated \texttt{calc} attribute.

In addition to an \texttt{atoms} attribute, \texttt{Calculator} objects also have a \texttt{parameters} attribute where calculator-specific settings are stored, as well as a \texttt{results} attribute, where the results of the calculation are stored --- just like in \texttt{ASE}.

\subsection{Scriptability}
\label{sec:scriptability}
Because \texttt{koopmans} is written in python, integrating it within a script is straightforward. For example, here is a script that runs the ozone calculation from Section~\ref{sec:ozone}:

\vspace{6pt}
\noindent\begin{minipage}{\columnwidth}%
\begin{minted}{python}
from ase import build
from koopmans.workflows import SinglepointWorkflow

# Create an Atoms object
atoms = build.molecule('O3', vacuum=5.0)

# Create a koopmans Workflow object
workflow = SinglepointWorkflow(atoms=atoms, ecutwfc = 65.0, nbnd = 10)

# Run the workflow
workflow.run()

# Fetch the IP and EA
results = workflow.calculations[-1].results
ip = -results['homo_energy']
ea = -results['lumo_energy']

# Print the IP and EA to screen
print(f' IP = {ip:.2f} eV')
print(f' EA = {ea:.2f} eV')

\end{minted}
\end{minipage}

\vspace{6pt}
Of course, printing the IP and EA to screen is of limited value --- in reality at this stage the user would then generate plots, or feed these results to another code.

Often a user will want to run workflows and analyse data separately --- for example, they might run their workflow on remote high performance computing resources, and then, days later, analyse the results on their laptop. To permit this, \texttt{koopmans} generates a \texttt{.kwf} file when a workflow is run. This file can be loaded into python in order to recover the \texttt{Workflow} python object. For example, we could perform exactly the same analysis on our previous ozone calculation by replacing lines 1-11 with

\vspace{6pt}
\noindent\begin{minipage}{\columnwidth}%
\begin{minted}{python}
from koopmans import io
workflow = io.read('ozone.kwf')

\end{minted}
\end{minipage}

\vspace{6pt}
\noindent where \texttt{ozone.kwf} has been generated by some previously completed \texttt{koopmans} calculation.

In the above, we used the \texttt{SinglepointWorkflow} for running a Koopmans workflow from start to finish. \texttt{koopmans} implements several other workflows that automate tasks that are useful when performing Koopmans calculations, such as convergence testing, standalone Wannierization, and DFT calculations.

\subsection{Code quality and testing}
\texttt{koopmans} contains an extensive test suite implemented with \texttt{pytest} \cite{Krekel2004}. It also has typing annotations which allow it to be statically typechecked using \texttt{mypy}.

%% file: supercell_workflow.tex
\begin{tikzpicture}[font=\tiny, x=3.5cm, y=1cm]
    \begin{pgfonlayer}{background}
        % \node[fit= (KC init) (empty label) (filled label) (sc loop 1) (converged label), fill=seaborn_bg_grey, inner sep=0.5cm] (calculating screening) {};
        % \node [dummy, above=0cm of calculating screening, font=\sffamily]{Calculating screening parameters};
        \fill [seaborn_bg_grey_dark] (-2.1,-0.8) rectangle (1.3,4.6);
        \node at (-2.1, 4.6) [default_text] {\large Initialisation (molecules)};
        \fill [seaborn_bg_grey_dark] (-2.1,-4.6) rectangle (1.3,-1.2);
        \node at (-2.1, -1.2) [default_text] {\large Initialisation (solids)};
        \fill [seaborn_bg_grey_dark] (1.4,-4.6) rectangle (6.9,4.6);
        \node at (1.4, 4.6) [default_text] {\large Calculating screening parameters};
        \fill [seaborn_bg_grey_dark] (7,-4.6) rectangle (8,4.6);
        \node at (7, 4.6) [default_text, text width=3.5cm] {\large Final calculation};
        \fill [seaborn_bg_grey_dark] (8.1,-4.6) rectangle (9.1,4.6);
        \node at (8.1, 4.6) [default_text, text width=3.5cm] {\large Postprocessing (solids)};
    \end{pgfonlayer}

    % Key
    \node at (6.85, 5.25) [cp, text width=1.4cm, minimum height=0.7cm] {\texttt{kcp}};
    \node at (7.35, 5.25) [pw, text width=1.4cm, minimum height=0.7cm] {\texttt{pw}};
    \node at (7.85, 5.25) [wannier90, text width=1.4cm, minimum height=0.7cm] {\texttt{w90}};
    \node at (8.35, 5.25) [bespoke, text width=1.4cm, minimum height=0.7cm, font=\tiny\sffamily] {bespoke code};
    \node at (8.85, 5.25) [observable, text width=1.4cm, minimum height=0.7cm, font=\tiny\sffamily] {quantity of interest};

    % Initialisation
    % Option 1
    \node at (-1, 2.9) [cp] (DFT init) {DFT};
    \node at (0, 2.9) [cp] (PZ innerloop) {\scriptsize PZ unitary rotation};
    \path [line] (DFT init) -- (PZ innerloop);

    % OR
    \node at (-0.5, 1.9) [default] (or) {or};

    % Option 2
    \node at (-0.5, 0.9) [cp] (PZ init) {PZ};

    % Solids
    \node at (-1.5, -2.9) [pw] (pw DFT init) {\scriptsize DFT (primitive cell)};
    \node at (-0.5, -2.9) [wannier90] (wannierize) {wannierize};
    \node at (0.5, -2.9) [bespoke] (unfold) {fold to supercell};
    \path [line] (pw DFT init) -- (wannierize);
    \path [line] (wannierize) -- (unfold);

    % Calculating screening parameters
    \node at (2, 0) [cp] (KC init) {$\alpha^n$KI/$\alpha^n$KIPZ};

    \path let
    \p1 = (PZ innerloop),
    \p2 = (unfold.east)
    in
    coordinate (dummy) at (\x2, \y1);
    \path let
    \p1 = (PZ init),
    \p2 = (unfold.east)
    in
    coordinate (dummy2) at (\x2, \y1);
    \path [line] (PZ innerloop) -- (dummy) to[-|-=0.3] ([yshift=2\myyshift]KC init.west);
    \path [line] (PZ init.east) -- ([xshift=-3.5\myyshift]dummy2) to[-|-=0.3] (KC init.west);
    \path [line] (unfold.east) to[-|-=0.3] ([yshift=-2\myyshift]KC init.west);

    % KI filled %%%%%%%%%%%%%%%%%%%%%%%%%%%%%%%%%%%%%%%%%%%%%%%%%%

    % calculations
    \node at (3, 3) [cp] (N-1_filled) {\scriptsize DFT/$\alpha^n$KIPZ ($N-1$)};
    % \node at (3, 2) [cp] (DFT_filled) {DFT};
    % \node at (3, -2) [cp] (DFT_empty) {DFT};
    \node at (3, -3) [cp] (N+1_empty) {\scriptsize DFT/$\alpha^n$KIPZ ($N+1$)};

    % \path [line] (KC init) to[-|-] (DFT_filled);
    \path [line] (KC init) to[-|-] (N-1_filled);
    % \path [line] (KC init) to[-|-] (DFT_empty);
    \path [line] (KC init) to[-|-] (N+1_empty);

    % results
    \node at (4, 3) [observable] (EN-1_filled) {$E_i(N-1)$};
    \node at (4, 2) [observable] (lambda0_filled) {$\lambda_{ii}(0,1)$};
    \node at (4, 1) [observable] (lambda_filled) {$\lambda_{ii}({\alpha^n},1)$};
    \node at (4, 0) [observable] (EN) {$E(N)$};
    \node at (4, -1) [observable] (lambda_empty) {$\lambda_{ii}(\alpha^n,0)$};
    \node at (4, -2) [observable] (lambda0_empty) {$\lambda_{ii}(0,0)$};
    \node at (4, -3) [observable] (EN+1_empty) {$E_i(N+1)$};

    \path [line] (KC init) -- (EN);
    \path [line] (KC init.east) to[-|-] (lambda_filled.west);
    \path [line] (KC init.east) to[-|-] (lambda_empty.west);
    \path [line] (KC init.east) to[-|-] (lambda0_filled.west);
    \path [line] (KC init.east) to[-|-] (lambda0_empty.west);

    % \path [line] (DFT_filled) -- (lambda0_filled);
    \path [line] (N-1_filled) -- (EN-1_filled);

    % \path [line] (DFT_empty) -- (lambda0_empty);
    \path [line] (N+1_empty) -- (EN+1_empty);

    % alpha parameters
    \node at (5, 1.5) [observable] (alpha filled) {$\alpha^{n+1}_{i \in \text{filled}}$};
    \node at (5, -1.5) [observable] (alpha empty) {$\alpha^{n+1}_{i \in \text{empty}}$};

    \path [line] (lambda_filled) to[-|-] (alpha filled);
    \path [line] ([yshift=\myyshift]EN.east) to[-|-] (alpha filled.west);
    \path [line] (lambda0_filled) to[-|-] (alpha filled);
    \path [line] (EN-1_filled) to[-|-] (alpha filled);

    \path [line] (lambda_empty) to[-|-] (alpha empty);
    \path [line] ([yshift=-\myyshift]EN.east) to[-|-] (alpha empty.west);
    \path [line] (lambda0_empty) to[-|-] (alpha empty);
    \path [line] (EN+1_empty) to[-|-] (alpha empty);

    % SC check
    \coordinate (sc check) at (6, 0);
    \path [headless_line] (alpha empty) to[-|-] (sc check);
    \path [headless_line] (alpha filled) to[-|-] (sc check);

    % SC loop
    \node [below=0.7cm of EN+1_empty] (sc loop y) {};
    \path let
    \p1 = (sc check),
    \p2 = (sc loop y)
    in
    coordinate (sc loop 1) at (\x1, \y2);
    \path [line] (sc check) -- node [midway, right, font=\sffamily] (not converged label) {\footnotesize $\{\alpha_i\}$ not converged} (sc loop 1) -| (KC init.south);

    % Final calc
    \node at (7.5, 0) [cp] (final KI) {$\alpha$KI/$\alpha$KIPZ};
    \path [line] (sc check) -- node [midway, above, font=\sffamily] (converged label) {\footnotesize $\{\alpha_i\}$ converged} (final KI);

    % Postproc
    \node at (8.6, 0) [bespoke] (upfold) {\scriptsize unfold to primitive};
    \path [line] (final KI) -- (upfold);

    % Boxes
    % Screening parameters
    \node [boxwhite, fit= (N-1_filled) (EN-1_filled),
        draw, dashed, fill opacity=0, inner sep=0.1cm](filled box){};
    \node [dummy, above=-0.05cm of filled box, font=\sffamily](filled label){\footnotesize one per filled orbital (index $i$)};
    \node [boxwhite, fit= (N+1_empty) (EN+1_empty),
        draw, dashed, fill opacity=0, inner sep=0.1cm](empty box){};
    \node [dummy, below=-0.05cm of empty box, font=\sffamily](empty label){\footnotesize one per empty orbital (index $i$)};

    % Labels
    \node at (DFT init.north east) [label] {a};
    \node at (PZ innerloop.north east) [label] {b};
    \node at (PZ init.north east) [label] {c};
    \node at (pw DFT init.north east) [label] {d};
    \node at (wannierize.north east) [label] {e};
    \node at (unfold.north east) [label] {f};
    \node at (KC init.north east) [label] {g};
    \node at (N-1_filled.north east) [label] {h};
    \node at (N+1_empty.north east) [label] {i};
    \node at (EN-1_filled.north east) [label] {j};
    \node at (lambda0_filled.north east) [label] {k};
    \node at (lambda_filled.north east) [label] {l};
    \node at (EN.north east) [label] {m};
    \node at (lambda_empty.north east) [label] {n};
    \node at (lambda0_empty.north east) [label] {o};
    \node at (EN+1_empty.north east) [label] {p};
    \node at (alpha filled.north east) [label] {q};
    \node at (alpha empty.north east) [label] {r};
    \node at (final KI.north east) [label] {s};
    \node at (upfold.north east) [label] {t};

\end{tikzpicture}

%% file: primitive_workflow.tex
\begin{tikzpicture}[font=\tiny, x=3.5cm, y=1cm]
   \begin{pgfonlayer}{background}
      \fill [seaborn_bg_grey_dark] (-2.2,-1) rectangle (1.2,1.5);
      \node at (-2.2, 1.5) [default_text] {\large Initialisation};
      \fill [seaborn_bg_grey_dark] (1.3,-1) rectangle (3.7,1.5);
      \node at (1.3, 1.5) [default_text] {\large Calculating screening parameters};
      \fill [seaborn_bg_grey_dark] (3.8,-1) rectangle (5.2,1.5);
      \node at (3.8, 1.5) [default_text, text width=3.5cm] {\large Final calculation};
      % \fill [seaborn_bg_grey_dark] (8.1,-1) rectangle (9.1,1.5);
      % \node at (8.1, 1.5) [default_text, text width=3.5cm] {\large Postprocessing};
   \end{pgfonlayer}

   % Key
   \node at (3.45, 2.15) [pw, text width=1.4cm, minimum height=0.7cm] {\texttt{pw\vphantom{0}}};
   \node at (3.95, 2.15) [wannier90, text width=1.4cm, minimum height=0.7cm] {\texttt{w90}};
   \node at (4.45, 2.15) [kcw, text width=1.4cm, minimum height=0.7cm] {\texttt{kcw}};
   \node at (4.95, 2.15) [observable, text width=1.4cm, minimum height=0.7cm, font=\tiny\sffamily] {quantity of interest};

   % Initialisation
   % Solids
   \node at (-1.5, 0) [pw] (pw DFT init) {DFT};
   \node at (-0.5, 0) [wannier90] (wannierize) {wannierize};
   \node at (0.5, 0) [kcw] (w2k) {Wannier function conversion};
   \path [line] (pw DFT init) -- (wannierize);
   \path [line] (wannierize) -- (w2k);

   % Calculating screening parameters
   \node at (2, 0) [kcw] (KC screen) {DFPT calculation};
   \node at (3, 0) [observable] (alphas) {$\{\alpha_i\}$};
   \path [line] (w2k) -- (KC screen);
   \path [line] (KC screen) -- (alphas);

   % Final calc
   \node at (4.5, 0) [kcw] (KC ham) {$\alpha_0$KI/$\alpha_0$KIPZ};
   \path [line] (alphas) -- (KC ham);

   % Labels
   \node at (pw DFT init.north east) [label] {u};
   \node at (wannierize.north east) [label] {v};
   \node at (w2k.north east) [label] {w};
   \node at (KC screen.north east) [label] {x};
   \node at (alphas.north east) [label] {y};
   \node at (KC ham.north east) [label] {z};
\end{tikzpicture}